\documentclass[english,notitlepages]{revtex4-1}
\usepackage[T1]{fontenc}
\usepackage[latin9]{inputenc}
\usepackage{geometry}
\usepackage{color}
\usepackage{babel}
\usepackage{refstyle}
\usepackage{amsmath}
\usepackage{graphicx}
\usepackage[unicode=true,pdfusetitle,
 bookmarks=true,bookmarksnumbered=false,bookmarksopen=false,
 breaklinks=false,pdfborder={0 0 0},pdfborderstyle={},backref=false,colorlinks=true]
 {hyperref}
\hypersetup{
 allcolors=blue}

\makeatletter
\def\pdfstartlink@attr{}
\makeatother

\usepackage[final]{changes}

\makeatletter

\AtBeginDocument{\providecommand\Figref[1]{\ref{Fig:#1}}}
\AtBeginDocument{\providecommand\eqref[1]{\ref{eq:#1}}}
\AtBeginDocument{\providecommand\Tabref[1]{\ref{Tab:#1}}}
\AtBeginDocument{\providecommand\Subsecref[1]{\ref{Subsec:#1}}}
\providecommand{\tabularnewline}{\\}
\RS@ifundefined{subsecref}
  {\newref{subsec}{name = \RSsectxt}}
  {}
\RS@ifundefined{thmref}
  {\def\RSthmtxt{theorem~}\newref{thm}{name = \RSthmtxt}}
  {}
\RS@ifundefined{lemref}
  {\def\RSlemtxt{lemma~}\newref{lem}{name = \RSlemtxt}}
  {}

\makeatother

\begin{document}
\title{Electron cyclotron drift instability and anomalous transport: two-fluid
moment theory and modeling}

\author{Liang Wang}
\email{lwang@pppl.gov}
\affiliation{Princeton Plasma Physics Laboratory, Princeton, New Jersey 08544, USA}
\affiliation{Department of Astrophysical Sciences, Princeton, University, Princeton New Jersey 08544, USA}

\author{Ammar Hakim}
\affiliation{Princeton Plasma Physics Laboratory, Princeton New Jersey 08544, USA}

\author{Bhuvana Srinivasan}
\affiliation{Department of Aerospace and Ocean Engineering, Virginia Tech, Blacksburg, Virginia 24060, USA}

\author{James Juno}
\affiliation{Princeton Plasma Physics Laboratory, Princeton New Jersey 08544, USA}

\begin{abstract}
In the presence of a strong electric field perpendicular to the magnetic field, the electron cross-field ($\rm{E \times B}$) flow relative to the unmagnetized ions can cause the so-called Electron Cyclotron Drift Instability (ECDI) due to resonances of the ion acoustic mode and the electron cyclotron harmonics. This occurs in, for example, collisionless shock ramps in space, and in $\rm{E \times B}$ discharge devices such as Hall thrusters. A prominent feature of ECDI is its capability to induce an electron flow parallel to the background E field at a speed greatly exceeding predictions by classical collision theory. Such anomalous transport is important due to its role in particle thermalization at space shocks, and in causing plasma flows towards the walls of $\rm{E \times B}$ devices, leading to unfavorable erosion and performance degradation, etc. 
The development of ECDI and anomalous transport is often considered requiring a fully kinetic treatment. In this work, however, we demonstrate that a reduced variant of this instability, and more importantly, the associated anomalous transport, can be treated self-consistently in a collisionless two-fluid framework without any adjustable collision parameter.  By treating both electron and ion species on an equal footing, the free energy due to the inter-species velocity shear allows the growth of an anomalous electron flow parallel to the background E field. We will first present linear analyses of the instability in the two-fluid five- and ten-moment models, and compare them against the fully-kinetic theory. At low temperatures, the two-fluid models predict the fastest-growing mode in good agreement with the kinetic result. Also, by including more ($>=10$) moments, secondary (and possibly higher) unstable branches can be recovered. The dependence of the instability on ion-to-electron mass ratio, plasma temperature, and background B field strength is also thoroughly explored. We then carry out direct numerical simulations of the cross-field setup using the five-moment model. The development of the instability, as well as the anomalous transport, is confirmed and in excellent agreement with theoretical predictions. The force balance properties are also studied using the five-moment simulation data.
This work casts new insights into the nature of ECDI and the associated anomalous transport and demonstrates the potential of the two-fluid moment model in efficient modeling of $\rm{E \times B}$ plasmas.

\end{abstract}

\maketitle

\section{Introduction}

In this work, we present two-fluid (electron and ion) investigations
of an electrostatic instability due to the electron ${\rm E\times B}$
drift relative to unmagnetized ions initially at rest, with the
wavevector perpendicular to the uniform background magnetic field.
In the fully kinetic description, this instability is often called
the electron cyclotron drift instability (ECDI) due to the coupling
between the ion acoustic wave and Doppler-shifted discrete electron cyclotron harmonics.

The research interest of ECDI dates back to the 1970s when Refs. \citep{Gary1970,Forslund1970a,Forslund1971a,Gary1970b,Wong1970a,Lampe1971a,Lampe1972a,Forslund1972a}
presented rather thorough kinetic analyses of this instability, motivated
primarily by laboratory observations of enhanced fluctuations in low-$\beta$,
collisionless plasma shocks perpendicular to a background magnetic
field. At the shock ramp, a fraction of the incoming ions are reflected
by the shock potential, picking up a fast drift relative to the incoming
electrons, exciting ECDI and other microinstabilities.
More recently, ECDI received revived interest
in the Earth's bow shocks due to the common observations of electron
Bernstein waves in association with ion acoustic waves \citep{Muschietti2006a,Muschietti2013b,Muschietti2017a,Wilson2010,Wilson2014,Breneman2013,Goodrich2018,Cohen2020a},
and was suggested to be a potentially important mechanism to allow
efficient electron bulk thermalization \citep{Wilson2014,Chen2018a}.

ECDI attracted significantly more attention in the Hall effect thruster
(HET) research community, stimulated by the continuing efforts to
develop electrically powered spacecraft propulsion~\citep{choueiri2001plasma,Goebel2008-pm,Boeuf2017}.
In the HET design,
a strong electric potential is applied between the anode at the closed
end of an annular ceramic channel, and a cathode external to the open
end of the channel. Propellant injected at the anode end are ionized
by electrons streaming from cathode and accelerated by the applied
electric field to produce thrust. To reduce the electron's axial mobility
towards the anode and prolong their residency time in the working
channel, a radial magnetic field is applied to magnetize the electrons
and confine them through the drift in the ${\rm E\times B}$ azimuthal
direction. In numerous studies, however, enhanced electron axial (that
is, parallel to the applied electric field) mobility is observed that
cannot be explained by classical diffusion due to electron-neutral
or electron-ion collisions. 
A number of explanations have since been
proposed to understand this anomalous electron transport~\citep{Boeuf2017,Taccogna2019},
with the
most promising one being an azimuthal instability, which is the topic
of this manuscript, the ECDI.

The role of ECDI in the HET context has been actively studied
through laboratory experiments and fully kinetic Particle-in-Cell
modeling, which could be computationally challenging for full-device
studies. Existing fluid and hybrid (fluid-electrons-kinetic-ions/neutrals)
modeling efforts have also been successful in producing useful HET operation
results, but primarily rely on adjustable parameters, for instance,
an adjustable, anomalous
collision frequency, to reproduce the observed ECDI characteristics
and anomalous electron transport. The conjecture that the enhanced
mobility does not manifest self-consistently in a fluid or hybrid
framework, is often implied. Thorough reviews on the HETs and the
numerical efforts to model their physics to different levels of complexity,
including ECDI and anomalous drift, can be found in \citep{Goebel2008-pm,Boeuf2017,Taccogna2019,Hara2019}.

In this work, we show how the ${\rm E\times B}$ electron
drift induce an azimuthal instability in a warm two-fluid (electron-ion)
high-moment description without any adjustable parameters, and further,
leads to anomalous axial transport of the electrons. This model treats
all species, critically, the electrons, in the same manner by evolving
their velocity moments, namely number density, velocity, and pressure (and possibly more).
The development of an azimuthal instability in
this framework is not surprising, due to the free energy available
from the velocity difference between electrons and ions.
In the cold plasma limit, it reduces to the magnetized Buneman instability \citep{Buneman1962,Smolyakov2016,Janhunen2018b}, therefore the instability itself is not a new finding of this manuscript.
The generation of axial transport
due to this instability is less evident, though, but can be shown to be a natural result of the Lorentz force
applied on the electrons, and, as we will show, is implied by the
eigenstructure of the ${\rm E\times B}$ instability. We shall also
see that, somewhat similar to the fully kinetic description, this
instability is due to the coupling between the ion acoustic mode and the Doppler-shifted  hybrid wave where the electron cyclotron dynamics play
a critical role. However, with only the lower-order velocity moments
taken into account, effects due to higher cyclotron harmonics are
lost, leading to less or no quantization and consequently greater
deviation from a fully kinetic description as the plasma temperature increases.

{
In this manuscript, we do not intend to suggest the 5-moment two-fluid model as a replacement for the fully-kinetic model, nor to report the discovery of a new fluid instability. Instead, the goals of this paper include
(1) to suggest the growth of anomalous transport in a purely fluid description without any collision, which was previously thought to mandate kinetic treatment or anomalous transport;
(2) to explore the nature and scaling of the electron drift instability in a finite-temperature two-fluid 5-moment (scalar pressure) plasma;
(3) to explore regimes when the fluid and kinetic prediction make a meaningful (not perfect), order-of-magnitude agreement;
(4) to show that with more velocity moments included in this two-fluid, high-moment framework, the model may capture higher electron cyclotron harmonics and unstable branches, which helps the model to achieve better agreement with the fully-kinetic treatment at higher temperatures.
}

This manuscript is outlined as follows. In Section. \ref{sec:Fluid-Linear-Theory},
we present the two-fluid high-moment model framework and the
linear theory of the ECDI in the 5-moment model, which assumes adiabatic Maxwellian
plasma species, as well as some results
using the 10-moment model which captures more kinetic physics through higher-order moments. In Section. \ref{sec:Numerical-Simulations},
we perform two-fluid 5-moment simulations of the ${\rm E\times{\rm B}}$
configuration and demonstrate the development of ECDI and anomalous
electron transport.
In Section \ref{sec:Comparing-with-Vlasov}, we compare 5-moment and fully-kinetic Vlasov-Poisson simulations using experimental parameters to demonstrate the capapbilities as well as limitation of the former in capturing ECDI and anomalous current.
We conclude in Section. \ref{sec:Discussions-and-Conclusions}
by summarizing the results and providing future motivation to use
high-moment models for cross-field instability studies.

\section{\label{sec:Fluid-Linear-Theory}Fluid Linear Theory}

A unique and interesting feature of the ECDI in the  fully-kinetic description is the discrete growth rates near wave numbers $k_m=m\Omega_{ce}/v_{\rm E \times B}$, where $\Omega_{ce}$ and $v_{\rm E \times B}$ are the electron cyclotron frequency and drift velocity, and $m$ is an integer mode numbers.
This is evident in the example in Figure \ref{fig:dr-compare}c, where the green curves represent a typical dispersion relation of ECDI in the fully-kinetic (Vlasov) description. As shown in Figure \ref{fig:dr-compare}c, the kinetic growth rate of ECDI peaks near integer multiples of $\Omega_{ce}/v_{\rm E \times B}$ due to the coupling of the ion acoustic wave and Doppler-shifted Bernstein harmonics at integral multiples of $\Omega_{ce}$.
At lower temperatures, however, the quantized
unstable branches expand and eventually form
a single unstable interval, as indicated by the
growth rates shown in Figure \ref{fig:dr-compare}(a-b).
In this section, we will present linear analysis of the ECDI in the two-fluid 5-moment and 10-moment theories to demonstrate 
their similarities and differences from a fully-kinetic description in describing the ECDI.

\subsection{The two-fluid 5-moment and 10-moment models}

A key concept in the multifluid high-moment model framework is the
equal treatment of all populations in the plasma, critically, the
electrons. In other words, the electron flows, inertial, and thermal
effects are self-consistently evolved instead of inferred from assumptions
like quasi-neutrality, etc. In this work, we focus on the
5-moment model in this framework, which assumes the electron and ion
pressures to be isotropic, i.e., scalars \citep{Hakim2006},
\begin{equation}
\begin{aligned} & \frac{\partial\rho_{s}}{\partial t}+\nabla\cdot\left(\rho_{s}\mathbf{v}_{s}\right)=0,\\
 & \frac{\partial\left(\rho_{s}\mathbf{v}_{s}\right)}{\partial t}+\nabla p+\nabla\cdot\left(\rho_{s}\mathbf{v}_{s}\mathbf{v}_{s}\right)=n_{s}q_{s}\left(\mathbf{E}+\mathbf{v}_{s}\times\mathbf{B}\right),\\
 & \frac{\partial\mathcal{E}_{s}}{\partial t}+\nabla\cdot\left[\mathbf{v}_{s}\left(p_{s}+\mathcal{E}_{s}\right)\right]=n_{s}q_{s}\mathbf{v}_{s}\cdot\mathbf{E}.
\end{aligned}
\label{eq:5m-equations}
\end{equation}
Here, $\rho_{s}$, $\mathbf{v}_{s}$, $p_{s}$, and $\mathcal{E}_{s}=p_{s}/\left(\gamma_{{\rm gas}}-1\right)+\frac{1}{2}\rho_{s}v_{s}^{2}$
are the mass density, velocity, thermal pressure, and total energy
of the plasma population $s$.
For electrostatic problems (like ours),
the magnetic field is supplied as a background,
while the electric field is solved with the Poisson's
equation, coupling all plasma populations through
their charge densities \citep{Hakim2006,Wang2020a}.

In addition to the 5-moment model, in this work, we will also present
some results from the 10-moment model where plasma pressures are treated
as full tensors with potentially unequal diagonal and non-vanishing
off-diagonal elements \citep{Hakim2008}:
\begin{align*}
 & \frac{\partial n}{\partial t}+\frac{\partial}{\partial x_{j}}\left(nu_{j}\right)=0,\\
 & m\frac{\partial\left(nu_{i}\right)}{\partial t}+\frac{\partial\mathcal{P}_{ij}}{\partial x_{i}}=nq\left(E_{i}+\varepsilon_{ijk}u_{j}B_{k}\right),\\
 & \frac{\partial\mathcal{P}_{ij}}{\partial t}=nqu_{[i}E_{j]}+\frac{q}{m}\varepsilon_{[ikl}\mathcal{P}_{kj]}B_{l}.
\end{align*}
Here, we have neglected subscripts $s$ for simplicity, $
\mathcal{P}_{ij}=\int d\mathbf{v}mv_{i}v_{j}f\left(\mathbf{v}\right)$ is the stress tensor in the rest frame,
the square brackets around indices represent the minimal sum over permutations of free indices needed to yield completely symmetric tensors (for example, $u_{[i}E_{j]} = u_{i}E_{j} + u_{j}E_{i}$).
The 10-moment model retains
more kinetic effects resulting in one more electron harmonic which is particularly insightful for the ECDI physics. The 10-moment results will be used to
better demonstrate the capability of high-moment models in capturing ECDI, though this paper focuses on the five-moment results primarily.

\subsection{\label{subsec:dr-compare}Dispersion relations in the 5-moment, 10-moment,
and fully-kinetic Vlasov models}

\begin{figure*}

\begin{centering}
\includegraphics[width=0.495\textwidth]{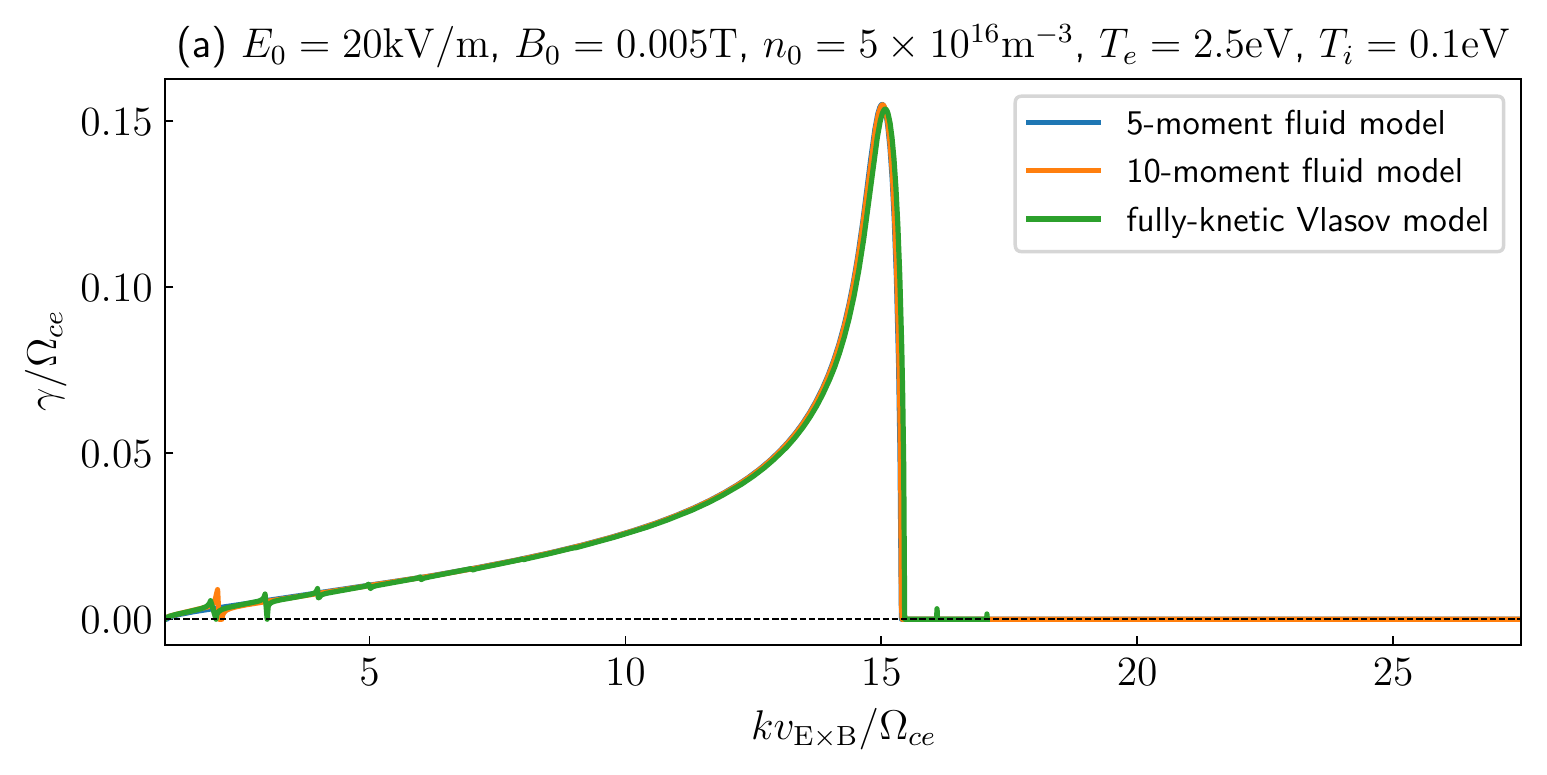}
\includegraphics[width=0.495\textwidth]{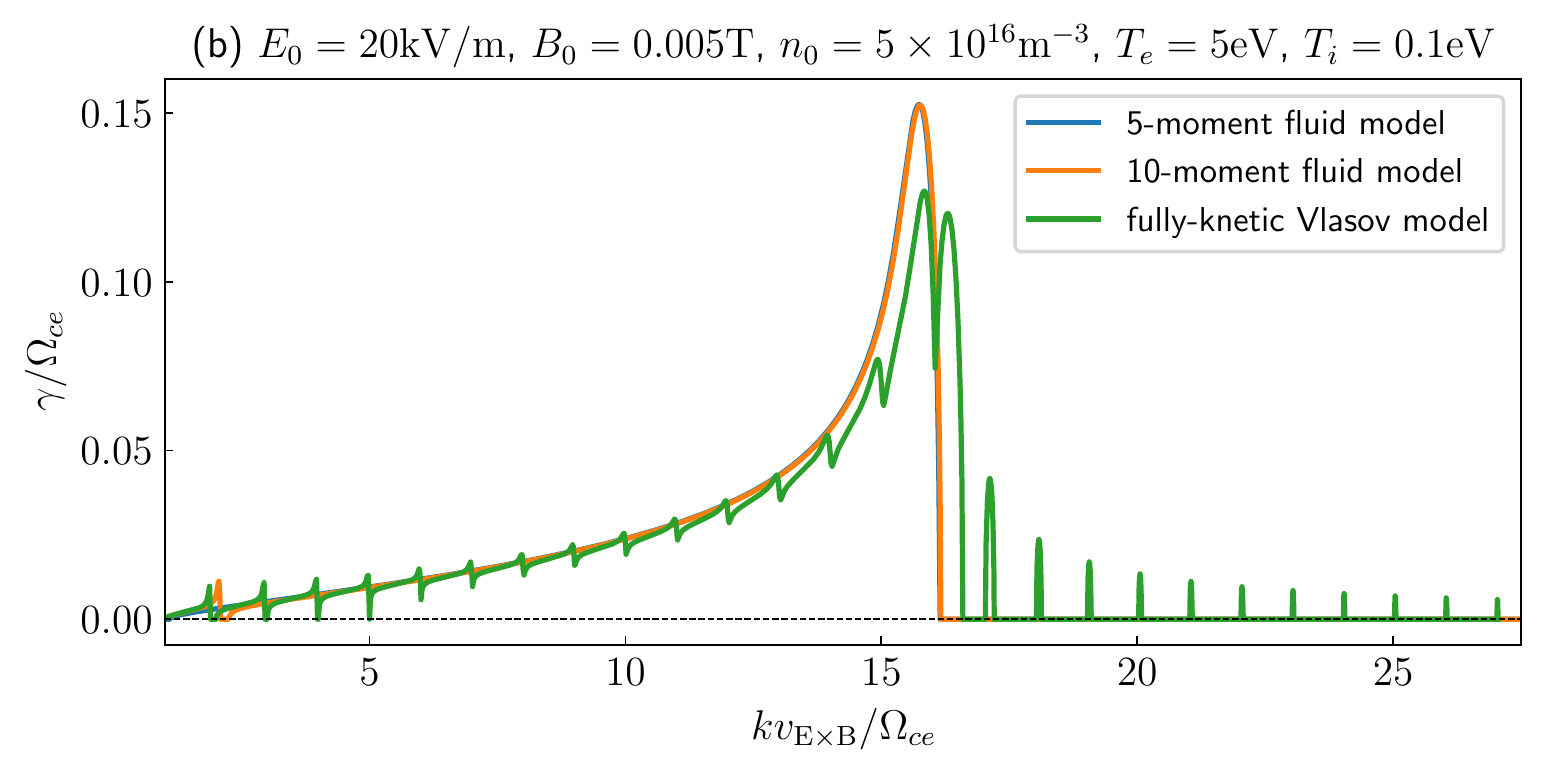}
\par\end{centering}
\begin{centering}
\includegraphics[width=0.495\textwidth]{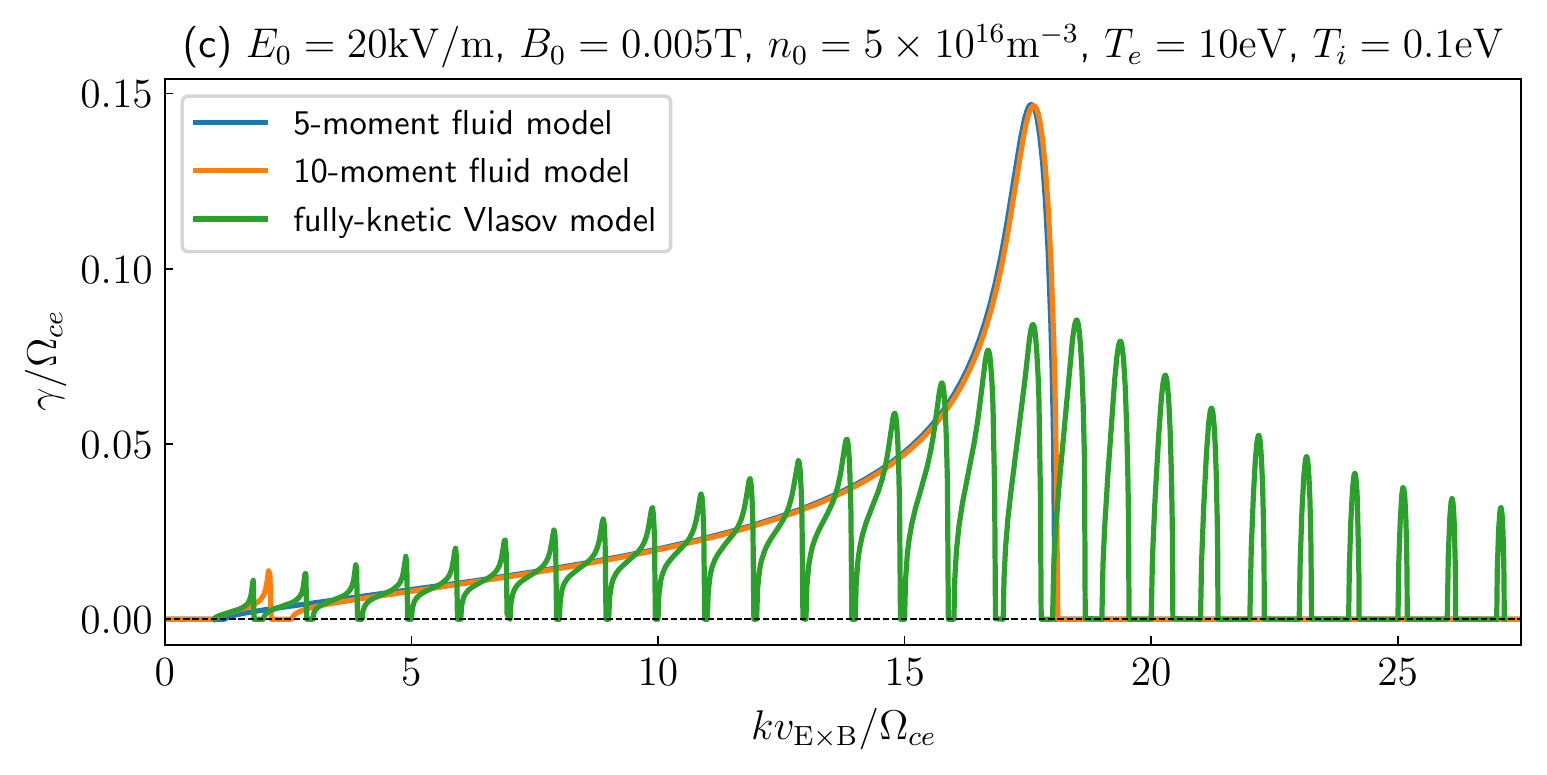}
\includegraphics[width=0.495\textwidth]{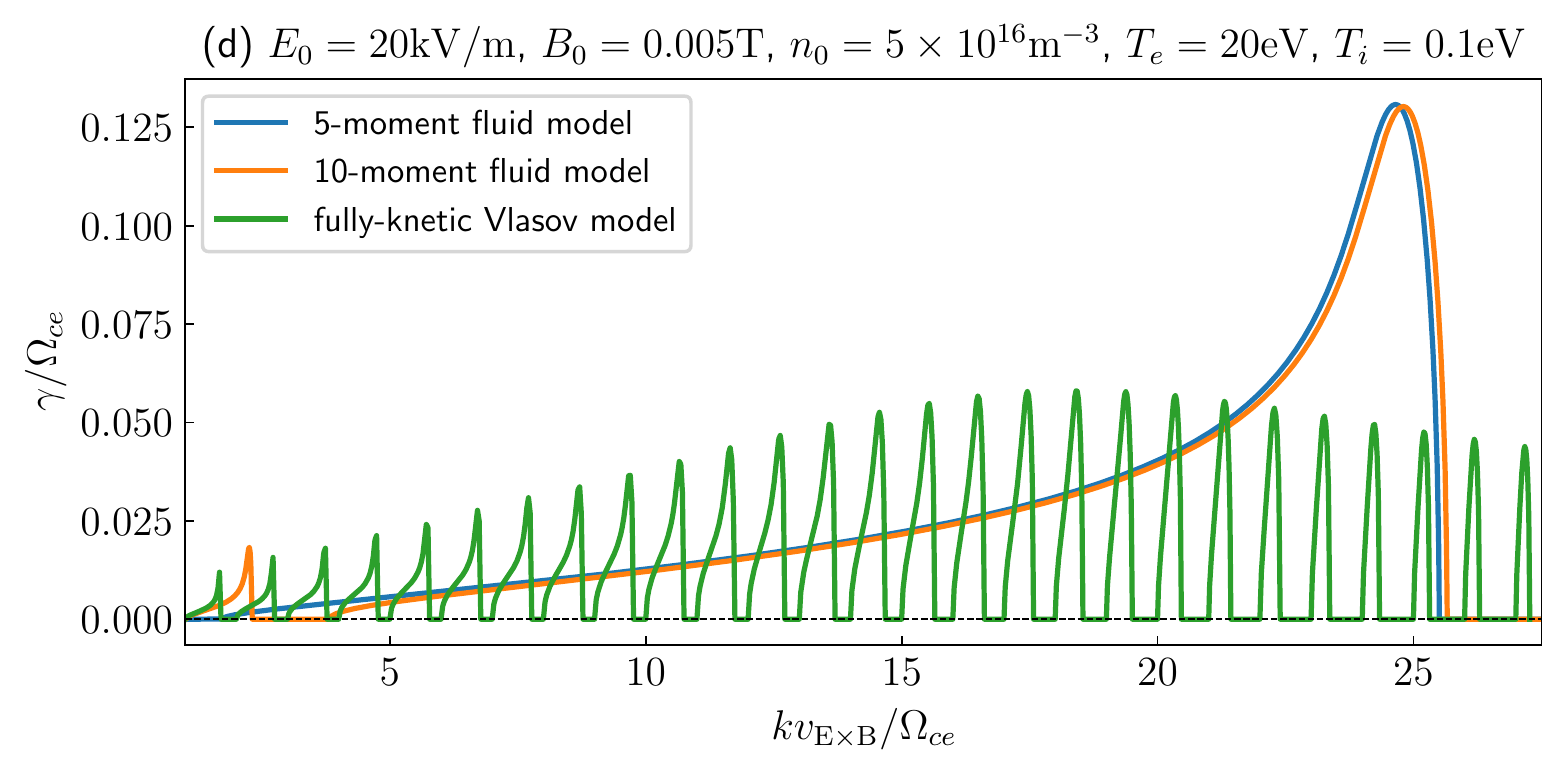}
\par\end{centering}

\begin{centering}
\includegraphics[width=0.99\textwidth]{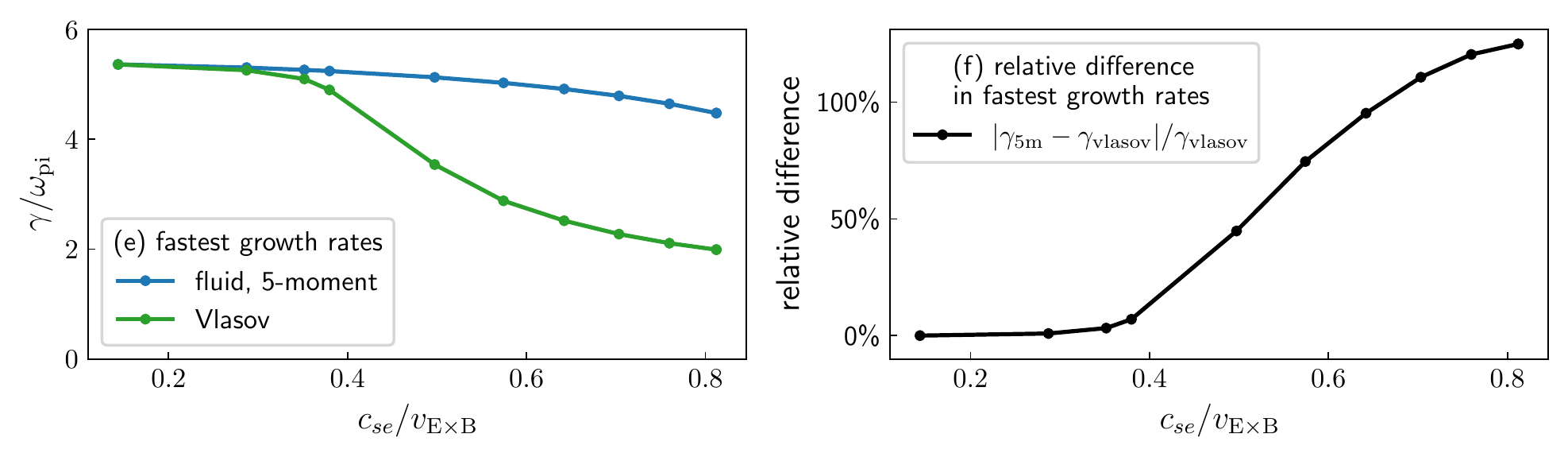}
\par\end{centering}

\caption{\label{fig:dr-compare}
Dispersion relations of the ECDI in fluid moment and fully-kinetic models at different electron/ion temperatures.
The parameters are 
$E_0=20~{\rm kV/m}$, $B_0=0.005~{\rm T}$, $n=5\times10^{16}~{\rm m^{-3}}$, $T_{i}=0.1~{\rm eV}$, and  $T_{e}$ values $1~{\rm eV}$, $5~{\rm eV}$, $10~{\rm eV}$, and $20~{\rm eV}$.
The 5-moment, 10-moment, and Vlasov dispersion relations are in blue, orange, and green, respectively.
Bottom panel (e) shows the growth rates of the fastest-growing mode (FGM) predicted by the 5-moment and the Vlasov models as a function of the ratio $c_{se}/v_{\rm E \times B}$. The bottom panel (f) shows the relative error between the 5-moment and Vlasov predictions of the FGM growth rates.
}
\end{figure*}

Consider 1D electrostatic modes in a 1D two-fluid plasma with
fully magnetized electrons and unmagnetized ions. The background magnetic
field $\mathbf{B}_{0}=B_{0}\hat{\mathbf{e}}_{z}$ is along $z$ and
the wavevector $\mathbf{k}=k\hat{\mathbf{e}}_{x}$ is along $x$,
perpendicular to $\mathbf{B}_{0}$. There exists a background electric
field $\mathbf{E}=E_{0}\hat{\mathbf{e}}_{y}$ and the fully magnetized
electrons flow at the ${\rm E\times B}$ drift velocity $\mathbf{v}_{{\rm E\times B}}=\mathbf{E}\times\mathbf{B}/B^{2}=v_{0}\hat{\mathbf{e}}_{x}$
along $x$.

The dispersion relation
in the 5-moment regime can be written as 
\begin{align}
1 & =\frac{\omega_{pi}^{2}}{\omega^{2}-k^{2}c_{si}^{2}}+\frac{\omega_{pe}^{2}}{\left(\omega-kv_{{\rm E\times B}}\right)^{2}-k^{2}c_{se}^{2}-\Omega_{ce}^{2}},\label{eq:ecdi-dr-5m-full}
\end{align}
where $c_{sj}=\sqrt{\gamma p_{j}/\rho_{j}}$, $\omega_{pj}$, and $\Omega_{cj}$ are the sound speed, plasma
and cyclotron frequencies  for
the species $j$, respectively. Compared to the well-known cold-plasma
Buneman instability, the formal difference here lies in the presence
of the electron cyclotron term, i.e., the role of the first electron-electron resonance. Therefore, the name ``electron cyclotron drift
instability'' is still proper for this fluid instability, though the
possible quantization due to higher electron cyclotron resonances
is missing and leads to greater deviation compared to the fully kinetic
counterpart when the plasma temperature goes up.
It's worth noting that the existence of drift instabilities in a magnetized two-fluid regime is not surprising, as indicated by \citet{Janhunen2018b} in the cold plasma limit. However, as we will see, this fluid instability has an important and surprising implication for anomalous transport, an important phenomenon in ${\rm E \times B}$ devices. This is what motivates our in-depth investigation of the origin and manifestation of the ECDI in the two-fluid moment model.

As mentioned earlier, the 10-moment model retains the full plasma
thermal pressure and one more electron resonance. Its dispersion relation
is more complex compared to the 5-moment counterpart and is not given here.
Instead, we numerically solve both the 5- and 10-moment dispersion
relations using a matrix-based algorithm \citep{Hakim2008,Xie2014,xenon}
to find the real and imaginary frequencies for any wavenumber and a set of
background physical parameters.

As an example, \Figref{dr-compare} show the growth rates from electrostatic 5-moment, 10-moment, and Vlasov dispersion relations for 
parameters used in the Particle-in-Cell benchmark work by Charoy et al.: discharge voltage $200~{\rm V}$, axial length $2.5~{\rm cm}$ radial magnetic field $B=0.001~{\rm T}$, number density $n=5\times10^{16}~{\rm m^{-3}}$, and electron and ion temperatures, $T_{e}=1~{\rm eV}$ and $T_{i}=0.5~{\rm eV}$.
Though the 5-moment and 10-moment results cannot obtain all the quantized
unstable branches that are seen in the kinetic dispersion, the moment models do provide reasonable estimation of fastest growth rate at lower $T_e$.
It is interesting to note, however,
that the 10-moment is able to capture a secondary unstable branch near $k=\Omega_{ce}/v_{\rm E \times B}$ due to its inclusion of the full pressure tensor.
In the future work, one may include even higher velocity moments recover higher electron cyclotron branches, giving better agreement with the Vlasov prediction.


The agreement between 5-moment and Vlasov predictions can be better understood through the scaling laws in panels (e) and (f) of \Figref{dr-compare} using the parameters above but with varying $T_e$. For these parameters, the relative error is below  $25\%$ when the electron sound speed, $c_{se}$, is less than about 0.5 times the drift speed $v_{\rm E \times B}$. Similar estimations hold for typical Hall thruster parameters. At higher temperatures, higher moments would be required for the moment fluid code to achieve good agreement with the fully kinetic model. This will be investigated in future studies. As a first step, this paper will focus on the analysis of the 5-moment model, which will serve as the foundation for future research using higher-moment models.

\subsection{Location of the fastest-growing mode}

In the long wavelength limit, the 5-moment ECDI dispersion relation
has two asymptotic solutions: an ion-acoustic-like wave,
\begin{equation}
1\approx\frac{\omega_{pi}^{2}}{\omega^{2}-k^{2}c_{si}^{2}}\Rightarrow\omega^{2}-k^{2}c_{si}^{2}\approx\omega_{pi}^{2},\label{eq:dr-ion_acoustic}
\end{equation}
and a Doppler-shifted ``hybrid'' wave, given by 
\[
1\approx\frac{\omega_{pe}^{2}}{\left(\omega-kv_{{\rm E\times B}}\right)^{2}-k^{2}c_{se}^{2}-\Omega_{ce}^{2}},
\]
or
\begin{equation}
\left(\omega-kv_{{\rm E\times B}}\right)^{2}-k^{2}c_{se}^{2}\approx\Omega_{ce}^{2}+\omega_{pe}^{2}.\label{eq:dr-doppler_hybrid}
\end{equation}

The unstable region of the 5-moment ECDI on the $k$-$\omega$ graph
is near the crossings of the two waves. Neglecting
the $\omega_{pi}$ terms for simplicity and equating the two asymptotic
dispersion relations, we find the wavenumber of the fastest-growing
mode (FGM) in the $\omega>0$ and $k>0$ quadrant,
\begin{equation}
k_{{\rm FGM}}\approx\sqrt{\frac{\Omega_{ce}^{2}+\omega_{pe}^{2}}{\left(kv_{{\rm E\times B}}-c_{si}\right)^{2}-c_{se}^{2}}},\label{eq:k_fgm}
\end{equation}
and the associated real-frequency 
\begin{equation}
\omega_{{\rm FGM}}\approx\sqrt{k_{{\rm FGM}}^{2}c_{si}^{2}+\omega_{pi}^{2}}.\label{eq:wr_fgm}
\end{equation}

\begin{figure*}
\begin{centering}
\begin{minipage}[c]{0.75\textwidth}%
\begin{center}
\includegraphics[width=1\textwidth]{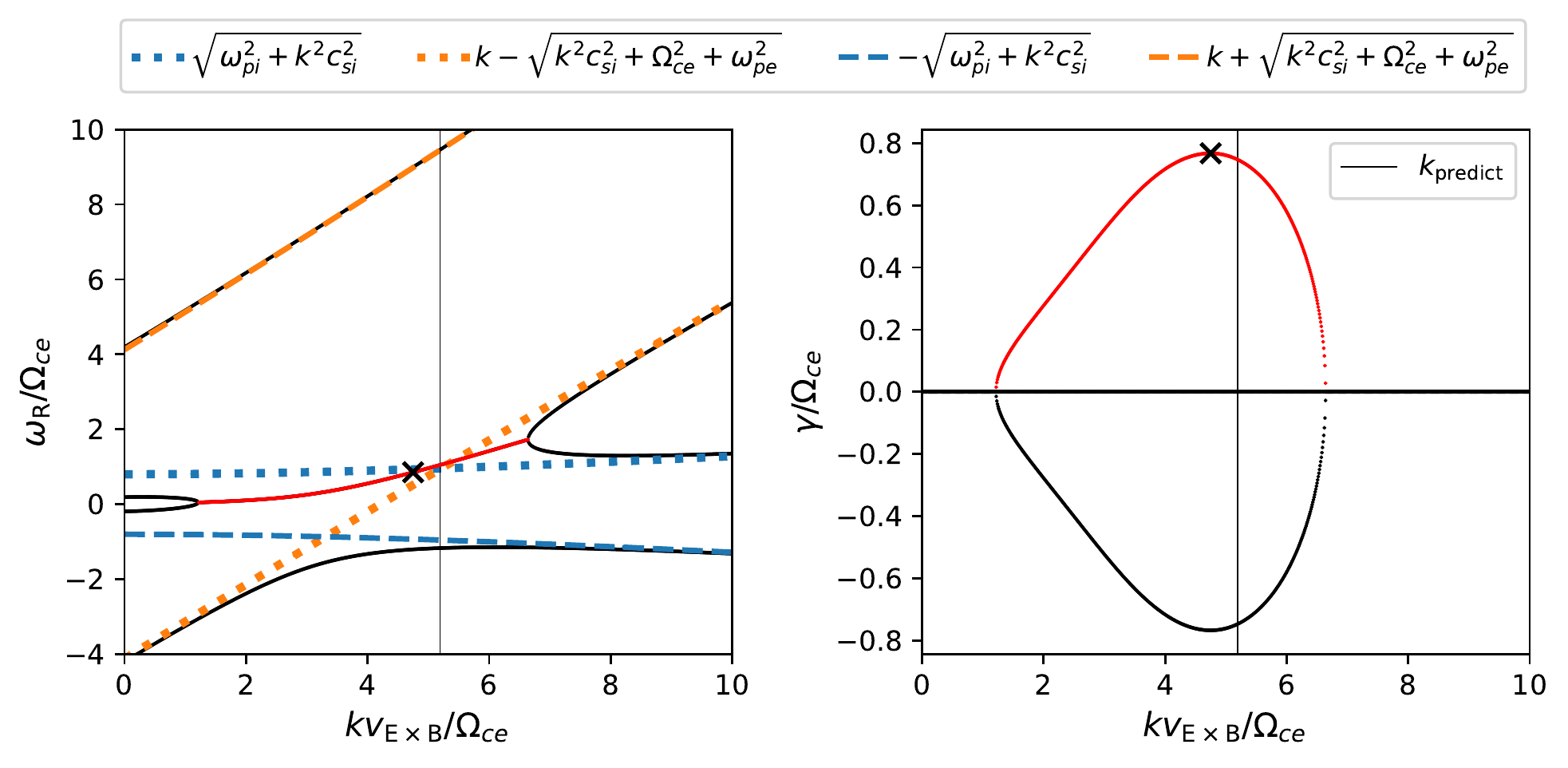}
\par\end{center}%
\end{minipage}
\par\end{centering}
\caption{\label{fig:fgm-crossing-ion_acoustic-hybrid}Predicting the location
of the fastest-growing mode using the crossing of the ion-acoustic
wave (\ref{eq:dr-ion_acoustic}) and the Doppler-shifted hybrid wave
(\ref{eq:dr-doppler_hybrid}). The left and right panels are the $k$-$\omega{\rm R}$ and $k$-$\gamma$ dispersion relation plots, respectively. For better visualization, simplified parameters are employed here: $m_{i}/m_{e}=25$,
$\omega_{pe}/\Omega_{ce}=4$, $c_{se}/v_{{\rm E\times B}}=0.2$, $c_{si}/v_{{\rm E\times B}}=0.1$.
The solid curves are the actual dispersion relation: the red curves
mark the branch with appreciable growth, while the black curves are
the rest of the full dispersion relation. The dotted and dashed lines
are asymptotic solutions at the large $k$ limit: the blue lines are
the ion-acoustic solution (\ref{eq:dr-ion_acoustic}), the orange
lines are the Doppler-shifted hybrid solution (\ref{eq:dr-doppler_hybrid}).
The $\boldsymbol{\times}$ symbol marks the actual location of the
fastest-growing mode in the $k>0$, $\omega>0$ quadrant. The thin,
black vertical lines mark the crossing of the dotted blue and orange
asymptotic lines, i.e., the $k_{{\rm FGM}}$ given by equation \ref{eq:k_fgm}
as a prediction of the location of the fastest-growing mode.}
\end{figure*}

The agreement between the numerically found (black crosses) and predicted (thin vertical line) locations of the fastest-growing
mode is shown in \Figref{fgm-crossing-ion_acoustic-hybrid} for a
set of artificial parameters listed in the figure caption. 
For a wide range of parameters, $k_{{\rm FGM}}$
and $\omega_{{\rm FGM}}$ give good prediction of the fastest-growing
mode's wavenumber and real-frequency, respectively.

\subsection{Dependence of the fastest growth rate on characteristic parameters}

The 5-moment ECDI dispersion relation (\ref{eq:ecdi-dr-5m-full})
can be written in the dimensionless form,
\begin{equation}
\frac{1}{r}=\frac{1}{m}\frac{1}{\tilde{\omega}^{2}-\tilde{k}^{2}c_{i}^{2}}+\frac{1}{\left(\tilde{\omega}-\tilde{k}\right)^{2}-\tilde{k}^{2}c_{e}^{2}-1},\label{eq:ecdi-dr-5m-dimensionless}
\end{equation}
where $\tilde{\omega}\equiv\omega/\Omega_{ce}$, $\tilde{k}\equiv kv_{{\rm {\rm E\times B}}}/\Omega_{ce}$,
$m\equiv m_{i}/m_{e}$, $r=\omega_{pe}/\Omega_{ce}$, $c_{e}=c_{se}/v_{{\rm E\times B}}$,
$c_{i}=c_{si}/v_{{\rm E\times B}}$. In other words, the
5-moment ECDI dispersion
relation is characterized by the four parameters: $m$, $r$, $c_{e}$,
$c_{i}$. It is thus useful to further understand how the fastest-growing
mode scales with these parameters. To this end, we start from the
baseline parameters $m=400$, $c_{e}=0.2$, and $c_{i}=0.02$, $r=10$,
and then vary them in isolation to understand how the ECDI mode growth scales.
{Note that these parameters are chosen not to match experiments but to amplify the effects of each parameter and make the scaling studies below more clear.}
Note that it is possible to further develop the analytical form of the
dispersion relation in various asymptotic limits but those will be
left for future work.

\begin{figure}
\begin{centering}
\begin{minipage}[c]{0.49\textwidth}%
\begin{center}
\includegraphics[width=1\textwidth]{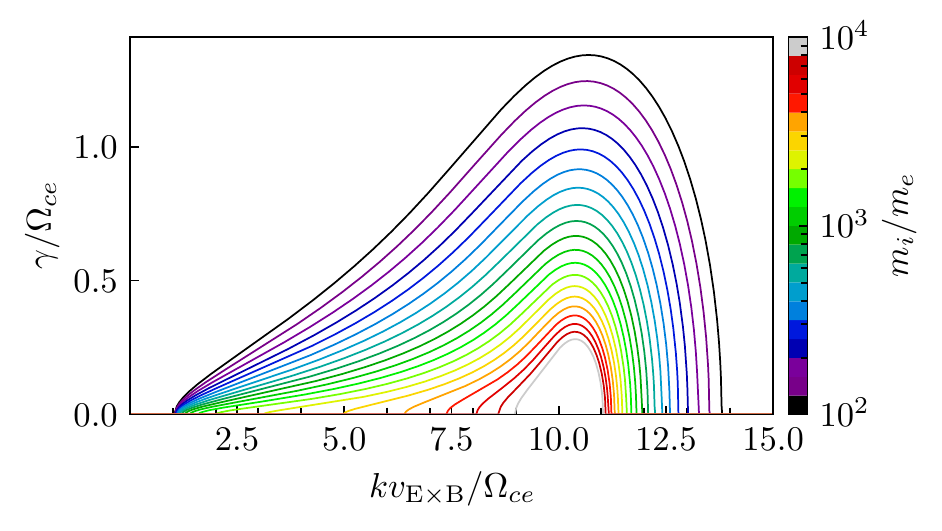}
\par\end{center}%
\end{minipage}
\par\end{centering}
\caption{\label{fig:scaling-mi}The dependence of the growth rate on the ion-to-electron
mass ratio, $m_{i}/m_{e}$. Different curves represent dispersion
relations due to a vast range of $m_{i}/m_{e}$ values.
All cases have identical
values for the following parameters: $\omega_{pe}/\Omega_{ce}=10$,
$\sqrt{\gamma p_{e}/nm_{e}}/v_{{\rm E\times B}}=0.2$, $\sqrt{\gamma p_{e}/nm_{e}}/v_{{\rm E\times B}}=0.02$.}
\end{figure}

\Figref{scaling-mi} shows the dependence of the growth rate on the
ion-to-electron mass ratio, $m_{i}/m_{e}$. It is clear to see that
the location of the fastest-growing mode does not change significantly
across a vast range of $m_{i}/m_{e}$ ratio. The maximum growth rate,
however, drops substantially as the mass ratio increases. The range of unstable
wavenumbers also shrinks as the ion-to-electron mass ratio increases, stacking the dispersion
relation curves. Another notable observation is that the unstable
range has a lower limit at $k\sim\Omega_{ce}/v_{{\rm E\times B}}$. 
The strong dependence of ECDI on ion-to-electron 
mass ratio indicates a critical difference in
the role of this instability in
the space plasmas, dominated by light ions like hydrogen and oxygen, versus HET plasmas, dominated by heavy ions like xenon and sometimes krypton.

\begin{figure*}
\begin{centering}
\begin{minipage}[c]{0.55\textwidth}%
\begin{center}
\includegraphics[width=1\textwidth]{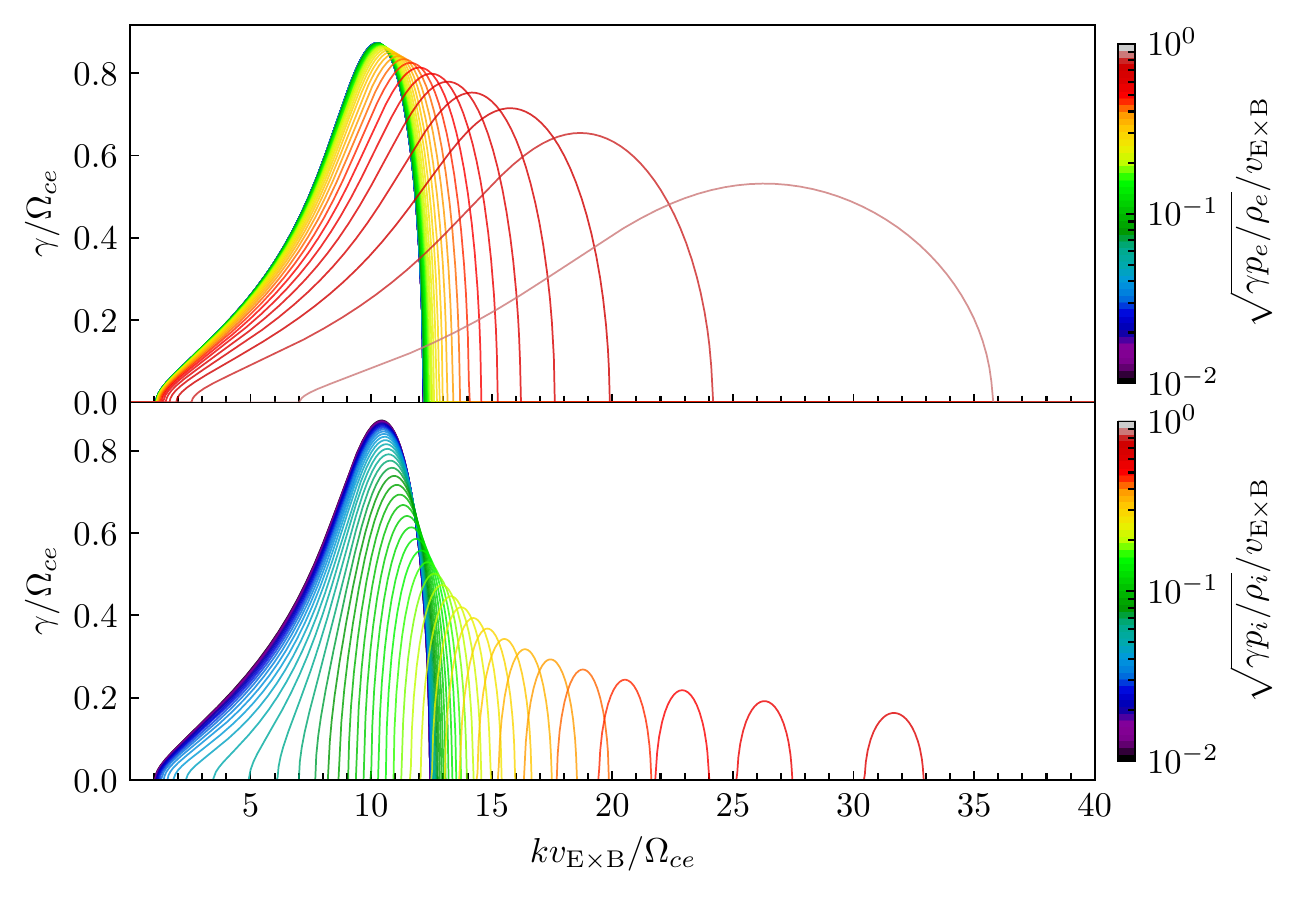}
\par\end{center}%
\end{minipage}%
\begin{minipage}[c]{0.45\textwidth}%
\begin{center}
\includegraphics[width=1\textwidth]{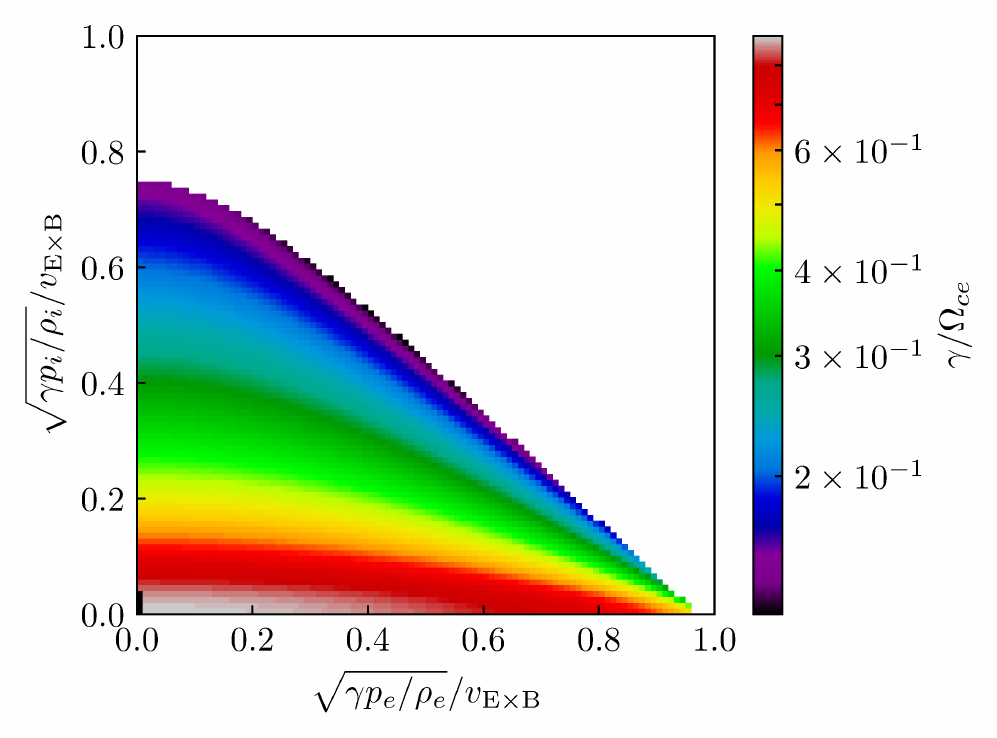}
\par\end{center}%
\end{minipage}
\par\end{centering}
\caption{\label{fig:scaling-temperature}The dependence of the growth rate
on the electron and ion sound speeds, and thus indirectly, their temperatures.
\textbf{Left two panels}: Different curves represent dispersion relations
due to different $\sqrt{\gamma p_{e}/nm_{e}}/v_{{\rm E\times B}}$
(left top) and $\sqrt{\gamma p_{e}/nm_{e}}/v_{{\rm E\times B}}$ (left
bottom) values. All cases have identical values for $m_{i}/m_{e}=400$
and $\omega_{pe}/\Omega_{ce}=10$. \textbf{Right panel}: The fastest-growing-mode
growth rates as a function of $\sqrt{\gamma p_{e}/nm_{e}}/v_{{\rm E\times B}}$
and $\sqrt{\gamma p_{e}/nm_{e}}/v_{{\rm E\times B}}$, within the
wavenumber interval $0<k<40v_{{\rm E}\times B}/\Omega_{ce}$.}
\end{figure*}

\Figref{scaling-temperature} demonstrates how the growth rate depends
on electron and ion sound speeds, and thus indirectly, the species temperatures.
From the left two panels, higher plasma temperature moves the interval
with nonzero growth towards greater wavenumbers and small growth rate. A more subtle
observation is that the growth rate drops slower with the electron
temperature than with the ion temperature. This is evident in the
right panel of \Figref{scaling-temperature}, which shows the maximum
growth rate over a matrix of $c_{e}$ and $c_{i}$ values. In this figure, the
gradient along the (vertical) ion axis is much greater than that along
the (horizontal) electron axis.

\begin{figure*}
\begin{centering}
\begin{minipage}[c]{0.49\textwidth}%
\begin{center}
\includegraphics[width=1\textwidth]{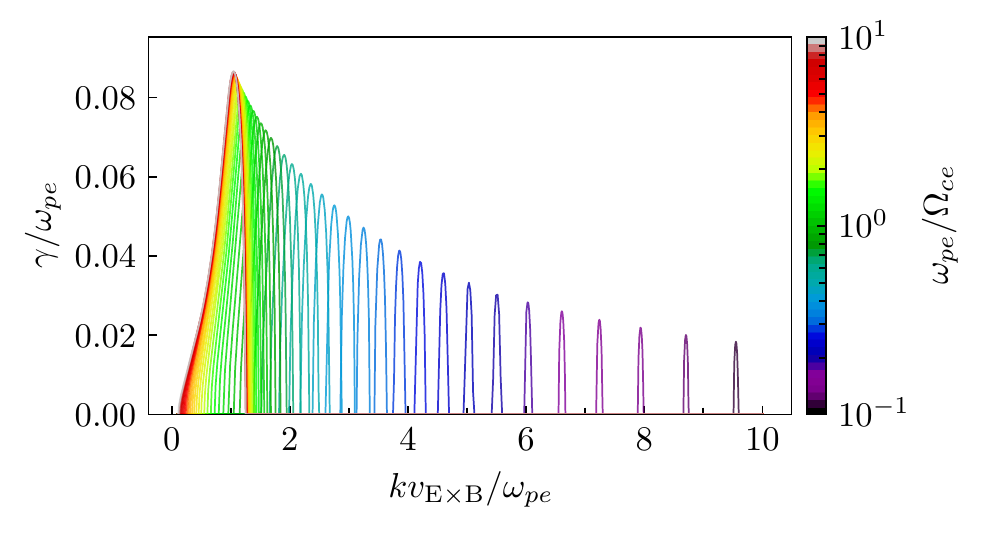}
\par\end{center}%
\end{minipage}
\par\end{centering}
\caption{\label{fig:scaling-B}The dependence of the growth rate on the background
magnetic field. Different curves represent different $\omega_{pe}/\Omega_{ce}$
values. All cases have identical values for the following parameters:
$m_{i}/m_{e}=400$, $\sqrt{\gamma p_{e}/nm_{e}}/v_{{\rm E\times B}}=0.2$,
$\sqrt{\gamma p_{e}/nm_{e}}/v_{{\rm E\times B}}=0.02$. 
The growth rates (vertical coordinates) are normalized by $\omega_{pe}$.}
\end{figure*}

\Figref{scaling-B} shows the dependence on the $\omega_{pe}/\Omega_{ce}$
ratio, and thus, indirectly, on the background magnetic field strength.
Here, the growth rates (vertical coordinates) are normalized by $\omega_{pe}$, which is fixed
across all cases.
The green curves have comparable
$\omega_{pe}$ and $\omega_{ce}$, while the blue curves have $\Omega_{ce}\gg\omega_{pe}$,
and red curves have $\Omega_{ce}\ll\omega_{pe}$.
In the strong magnetic
field limit (small $\omega_{pe}/\Omega_{ce}$ and blue curves),
the unstable range is very narrow and has large wavenumbers and low
growth rates. As the magnetic field weakens and its effect diminishes,
the dispersion relation changes less and less as it approaches the
Buneman instability in an unmagnetized plasma (the red curves).


\subsection{\label{subsec:eigenvector-of-simulaiton-parameters}Eigenvector and its indication of anomalous electron transport}

The matrix-based dispersion relation solver provides the eigenvectors associated with the eigenfrequencies. Here, we consider the parameters $m\equiv m_{i}/m_{e}=1836$,
$r=\omega_{pe}/\Omega_{ce}=5.07$, $c_{e}=c_{se}/v_{{\rm E\times B}}=0.3$,
$c_{si}/v_{{\rm E\times B}}=0.0022$.
Its fastest-growing mode occurs at $k\approx5.44322\,\Omega_{ce}/v_{{\rm E\times B}}$
with a growth rate $\gamma=0.26892\Omega_{ce}^{-1}$. Its normalized
eigenvector is listed in \Tabref{normalized-eigenvector}. Particularly,
the value for $v_{ye1}$ is non-trivial. In linear analysis, this
term and the $v_{xe1}$ perturbation stem from the Lorentz force due
to the background magnetic field. A direct consequence of the non-trivial
$v_{ye1}$ value is the development of appreciable anomalous axial
(i.e., along the applied electric field) transport of the electrons.
Direct numerical simulation of
anomalous electron transport is discussed in the next section.

\begin{table*}
\begin{centering}
\begin{tabular}{|c|c|c|c|}
\hline 
$n_{e1}$ & $v_{xe1}$ & $v_{ye1}$ & $v_{ze1}$\tabularnewline
\hline 
0.987 & 0.0181i & 0.0181i & 0\tabularnewline
\hline 
$n_{i1}$ & $v_{xi1}$ & $v_{yi1}$ & $v_{zi1}$\tabularnewline
\hline 
0.0636+0.109i & -0.000331+0.000666i & 0 & 0\tabularnewline
\hline 
\end{tabular}
\par\end{centering}
\caption{\label{tab:normalized-eigenvector}Normalized eigenvector for the
fastest-growing mode for parameters used by the simulation in \Secref{Numerical-Simulations}.}

\end{table*}

\section{\label{sec:Numerical-Simulations}Numerical Simulations Scanning Simplified Parameters}

In this section, we perform direct numerical simulations of the ECDI
by integrating the 5-moment equations (\ref{eq:5m-equations}), coupled
with the Poisson's equation for the electric field. The simulations are performed using the multi-moment solvers in the Princeton
code, Gkeyll \citep{Hakim2006,Hakim2008,Wang2020a}, that has been verified
extensively for a number of  plasma physics problems \citep{Wang2015,Ng2015,Ng2017,Wang2018b,Ng2018a,Ng2019,Dong2019b,TenBarge2019a,Ng2020, srinivasan2018role, cagas2017nonlinear}.
Similar models have also been implemented by other groups for various applications~\citep{Miller2016,meier2021development,Abgrall2014,AllmannRahn2018,allmann2021fluid,joncquieres201810,Laguna2019}.

\subsection{Simulation setup}

The simulation uses the 1D cross-field configuration, where the simulation domain is along $x$, the initial electric and magnetic fields are along $y$ and $z$, so that the initial electron drift is along $x$.
The wavenumber of the fastest-growing
mode is $k_{{\rm FGM}}=5.44\,\Omega_{ce}/v_{{\rm E\times B}}$
with a growth rate $\gamma_{{\rm FGM}}=0.27\Omega_{ce}^{-1}$.
Its normalized eigenvector is given in \Subsecref{eigenvector-of-simulaiton-parameters}. The
periodic simulation domain length is $L=10\lambda_{{\rm FGM}}=10\cdot2\pi/k_{{\rm FGM}}=11.54v_{{\rm E\times B}}/\Omega_{ce}$
and discretized with $N_{x}=1280$ cells. The CFL number is 0.95.
The gas gamma $\gamma_{{\rm gas}}$ is set to $3$.

Sinusoidal perturbations are applied to the electrical field $E_{x}$
so that the spectral energy $\left|E_{x}\left(k\right)\right|^{2}$
is evenly distributed across $kL/2\pi=0,1,...,32$. These modes then
compete with each other and, over time, the fastest-growing mode,
presumably the $k=k_{{\rm FGM}}$ one, dominates. The electron number
density $n_{e}$ is also perturbed to satisfy  Gauss's law initially.
Finally, the perturbation magnitudes are controlled by a parameter
$\delta=10^{-5}$ so that the mode $E_{x}\left(k_{m}\right)$ leads
to density fluctuation $\delta\cdot n_{e}/k_{m}L$ (and the same $\left|E_{x}\left(k_{m}\right)\right|$).

\subsection{Linear development}

\begin{figure*}
\begin{centering}
\begin{minipage}[c]{0.4\textwidth}%
\begin{center}
\includegraphics[width=1\textwidth]{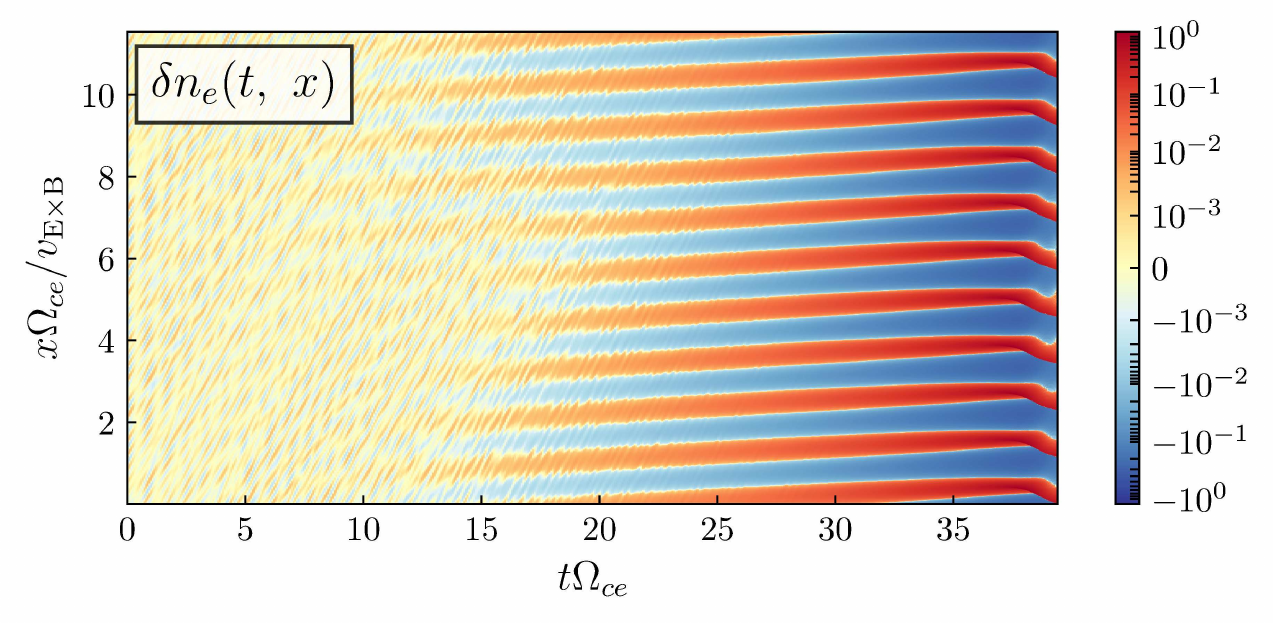}
\par\end{center}%
\end{minipage}%
\begin{minipage}[c]{0.4\textwidth}%
\begin{center}
\includegraphics[width=1\textwidth]{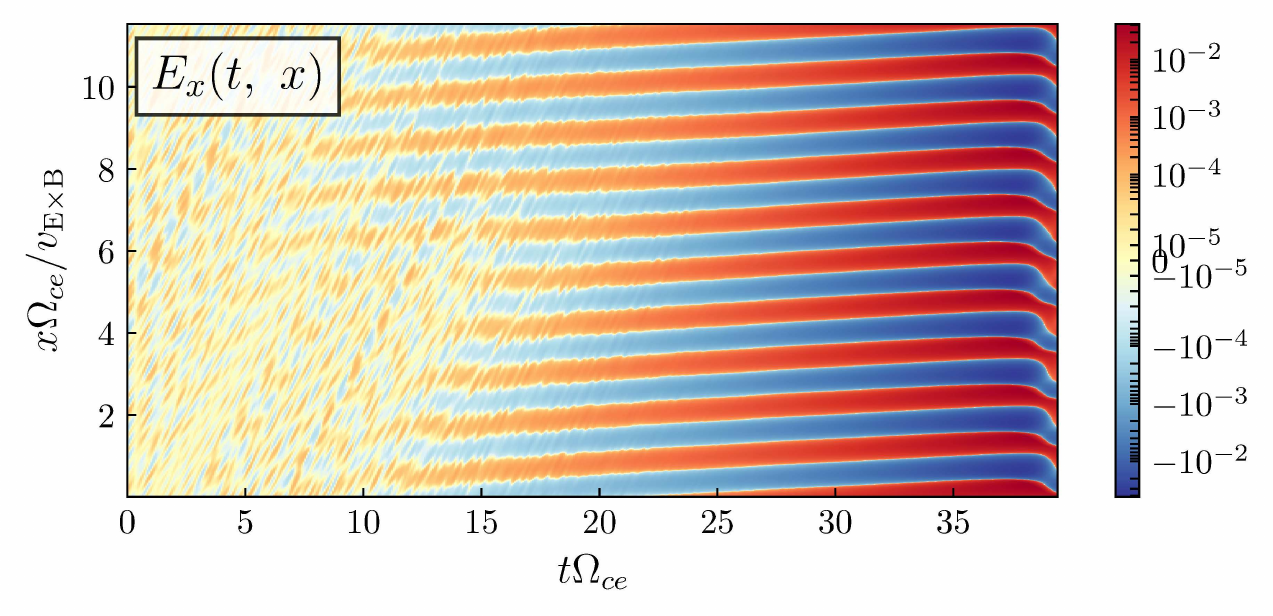}
\par\end{center}%
\end{minipage}
\par\end{centering}
\caption{\label{fig:time-evolution-ne-Ex}Time evolution of density fluctuation
$\delta n_{e}\equiv n_{e}-n_{e0}$ and azimuthal electric field $E_{x}$
in the numerical simulation.}
\end{figure*}

We first examine the development of electron density fluctuation and
azimuthal electric field shown in \Figref{time-evolution-ne-Ex}.
Both show clear dominance of a mode with a wavelength $\lambda=L/10$,
or $k=k_{{\rm FGM}}$ as predicted. Near the end of the simulation
at $t=40\Omega_{ce}^{-1}$, the linear development is saturated and
the simulation enters a nonlinear stage. In this paper, we focus on
the linear stage only.

An interesting observation is that the $n_{e}$ and $E_{x}$ are not
entirely out-of-phase; in other words, a nonzero average $\left\langle n_{e}E_{x}\right\rangle $
develops during the simulation. Consistent with Ref. \citep{Lafleur2016b}, this indicates a nonvanishing cross-field
electron mobility in the collisionless limit: $\mu_{\perp e}=-\left\langle n_{e}E_{{\rm azimuthal}}\right\rangle /n_{e}E_{{\rm axial}}B_{{\rm radial}}$
and enhances the electron anomalous transport. In the subsequent subsections,
we will study the anomalous transport in more detail.

\begin{figure*}
\begin{centering}
\begin{minipage}[c]{0.4\textwidth}%
\begin{center}
\includegraphics[width=1\textwidth]{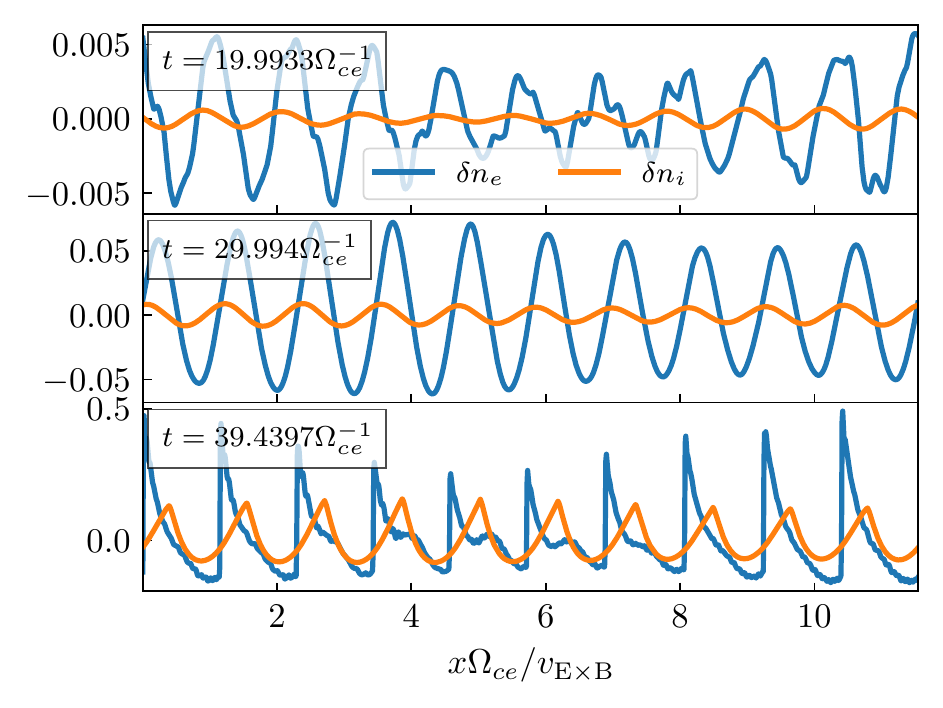}
\par\end{center}%
\end{minipage}%
\begin{minipage}[c]{0.4\textwidth}%
\begin{center}
\includegraphics[width=1\textwidth]{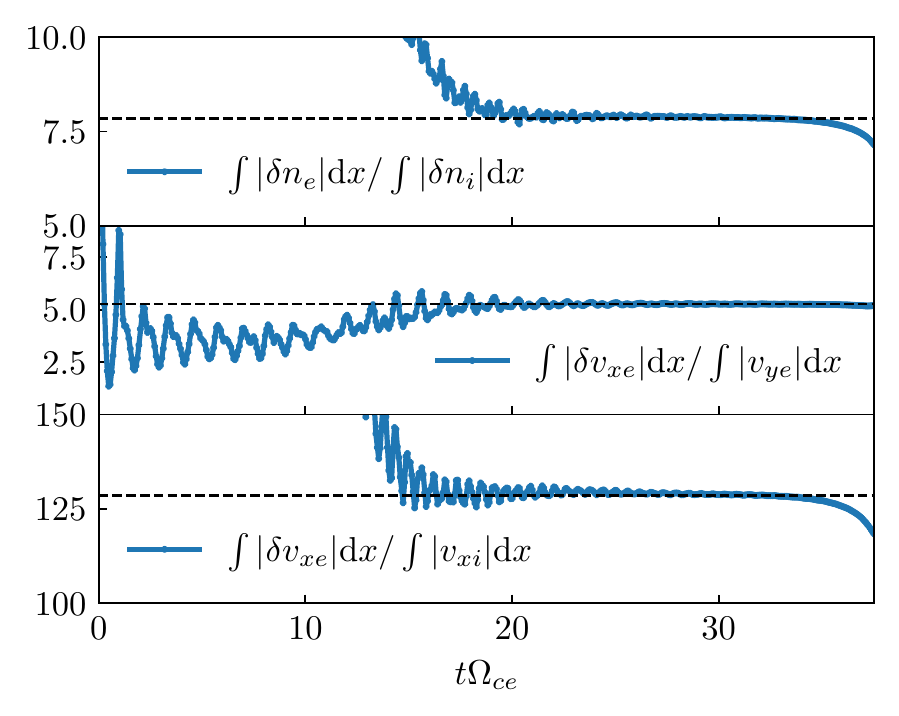}
\par\end{center}%
\end{minipage}
\par\end{centering}
\caption{\textbf{\label{fig:agreement-theory-eigenvector}Left}: Electron and
ion number density fluctuation in, from top to bottom, the earlier linear stage, later
linear stage, and early nonlinear stage.
The competing of various modes, the dominance of the fastest-growing mode, and
the steepening of the waves, are evident in these three stages.
\textbf{Right}: Time evolution
of ratios of spatially integrated fluctuations as they approach
values from linear-theory predictions of the fastest-growing mode.
The integrated quantities, from top to bottom, are $\int\left|n_{e}-n_{0}\right|{\rm d}x/\int\left|n_{i}-n_{0}\right|{\rm d}x$,
$\int\left|v_{xe}-v_{xe0}\right|{\rm d}x/\int\left|v_{xi}\right|{\rm d}x$,
and $\int\left|v_{xe}-v_{xe0}\right|{\rm d}x/\int\left|v_{ye}\right|{\rm d}x$.
Their expected values due to the fastest-growing mode from the linear
theory are marked by horizontal dashed lines, which are taken from
\Tabref{normalized-eigenvector}: $\left|n_{e1}\right|/\left|n_{i}\right|\approx7.8373$,
$\left|v_{xe1}\right|/\left|v_{ye1}\right|\approx5.27388$, $\left|v_{xe1}\right|/\left|v_{xi1}\right|\approx128.562$.}
\end{figure*}

Next, we compare the simulation results with the linear theory prediction.
The left panel of \Figref{agreement-theory-eigenvector} shows the
snapshots of the electron and ion density fluctuations at the early
linear stage, late linear stage, and early nonlinear stage. 
In the linear stage, $\delta n_{i}$ is only a small fraction of $\delta n_{i}$,
consistent with the egienvector prediction given in Table.~\ref{tab:normalized-eigenvector}.
Entering the nonlinear stage, the electron density profile becomes
highly spiky as the waves start to break. For a
more quantitative comparison to theory, the right panel of \Figref{agreement-theory-eigenvector}
shows the time evolution of different components of the eigenvectors using electron-to-ion ratios in blue curves,
along with their predicted values from the fastest-growing mode as
horizontal dashed lines. It is clear that beginning from about $t=16\Omega_{ce}^{-1}$,
the simulated ratios approaches the predicted values. Again, as predicted
by theory, a nonvanishing axial electron transport $\int v_{ye}{\rm d}x$
develops and leads to anomalous transport. Near the end of the simulation, where the evolution is nonlinear, the ratios begin to show deviations
from the linear prediction. These results provide excellent verification
of the linear development in our simulation.

\subsection{Anomalous electron transport}

\begin{figure*}
\begin{centering}
\begin{minipage}[c]{0.49\textwidth}%
\begin{center}
\includegraphics[width=1\textwidth]{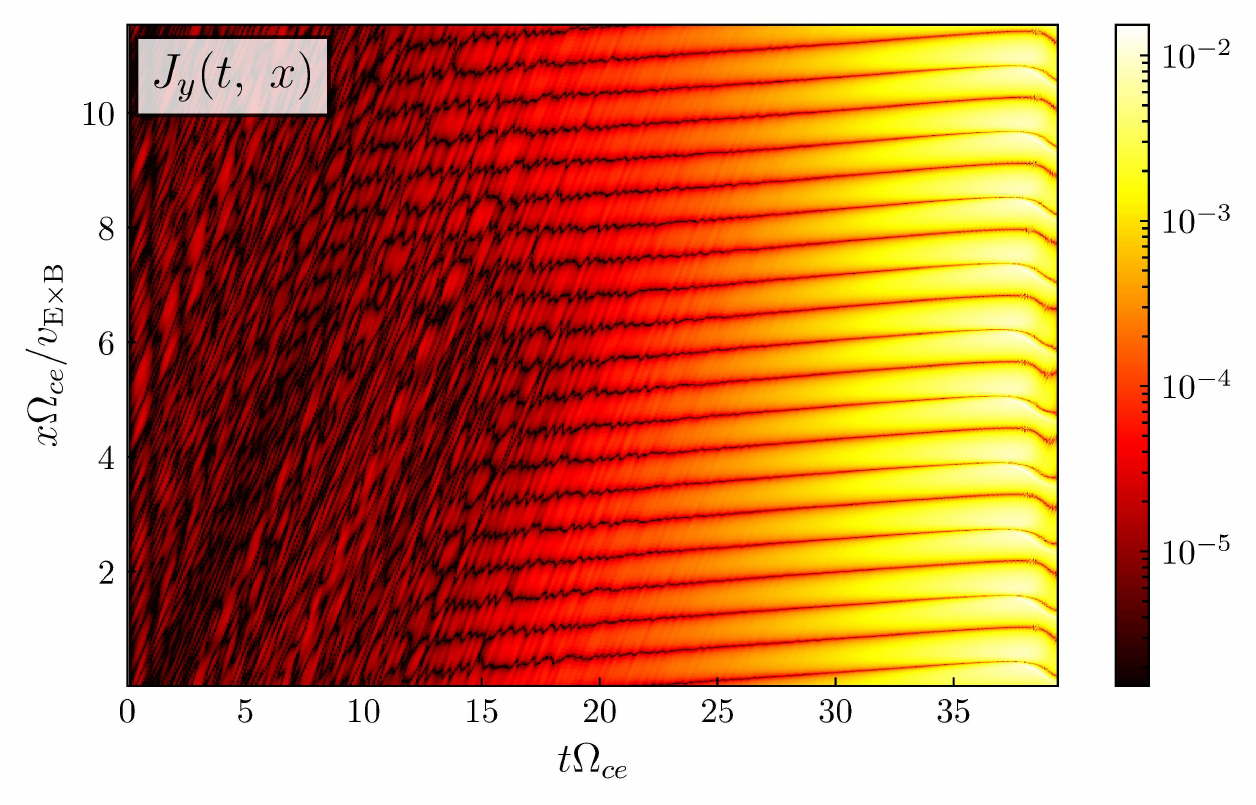}
\par\end{center}%
\end{minipage}%
\begin{minipage}[c]{0.49\textwidth}%
\begin{center}
\includegraphics[width=1\textwidth]{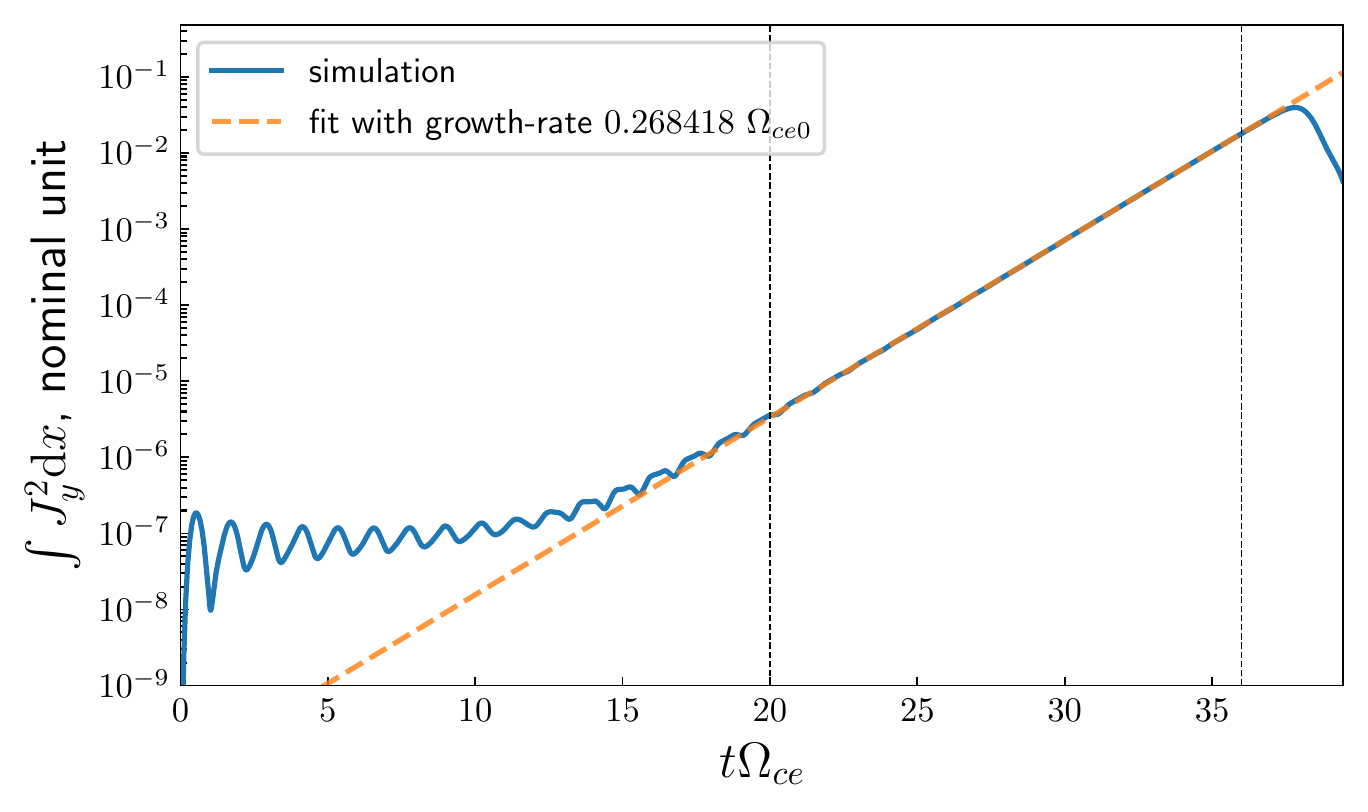}
\par\end{center}%
\end{minipage}
\par\end{centering}
\caption{\label{fig:time-evolution-anomalous-transport}Time evolution of anomalous
current in the numerical simulation. \textbf{Left}: Anomalous axial
current $J_{y}$ as a function of $\left(t,\,x\right)$. The horizontal
and vertical axes are time and $x$-coordinates, respectively. Note
that in this simulation $J_{y}$ is due to the anomalous transport
of electrons only since ions are unmagnetized and do not contribute
to $J_{y}$.\textbf{ Right}: Time evolution of the anomalous current
, $\int_{0}^{L}J_{y}{\rm d}x$, integrated over the entire domain
(blue curve) along with a linear fit (orange dashed line). The vertical
dashed lines mark the range where the fit is made.}
\end{figure*}

As mentioned earlier, the correlated fluctuations in $n_{e}$ and
$E_{x}$ indicate the existence of anomalous electron transport. The
left panel of \Figref{time-evolution-anomalous-transport} shows the
temporal-spatial profile of the anomalous electron current. A positive
net current develops in the linear stage at the predicted
wavelength. The right panel shows the growth of the integrated anomalous
current. In the early stage of the simulation, the competition between
modes of different wavelengths causes a wide spectrum of fluctuations. Beginning
from about $t=16\Omega_{ce}^{-1}$, one dominant mode arises
and its linear growth lasts about $20\Omega_{ce}^{-1}$, which is
determined by the initial perturbation level. Fitting the data between
the primary region of linear growth, at approximately $20<t\Omega_{ce}<37.5$, we find
a growth rate $\gamma=0.268418\Omega_{ce}$, which is in excellent
agreement with the theoretical prediction $0.26892\Omega_{ce}$.

\begin{figure*}
\begin{centering}
\begin{minipage}[c]{0.55\textwidth}%
\begin{center}
\includegraphics[width=1\textwidth]{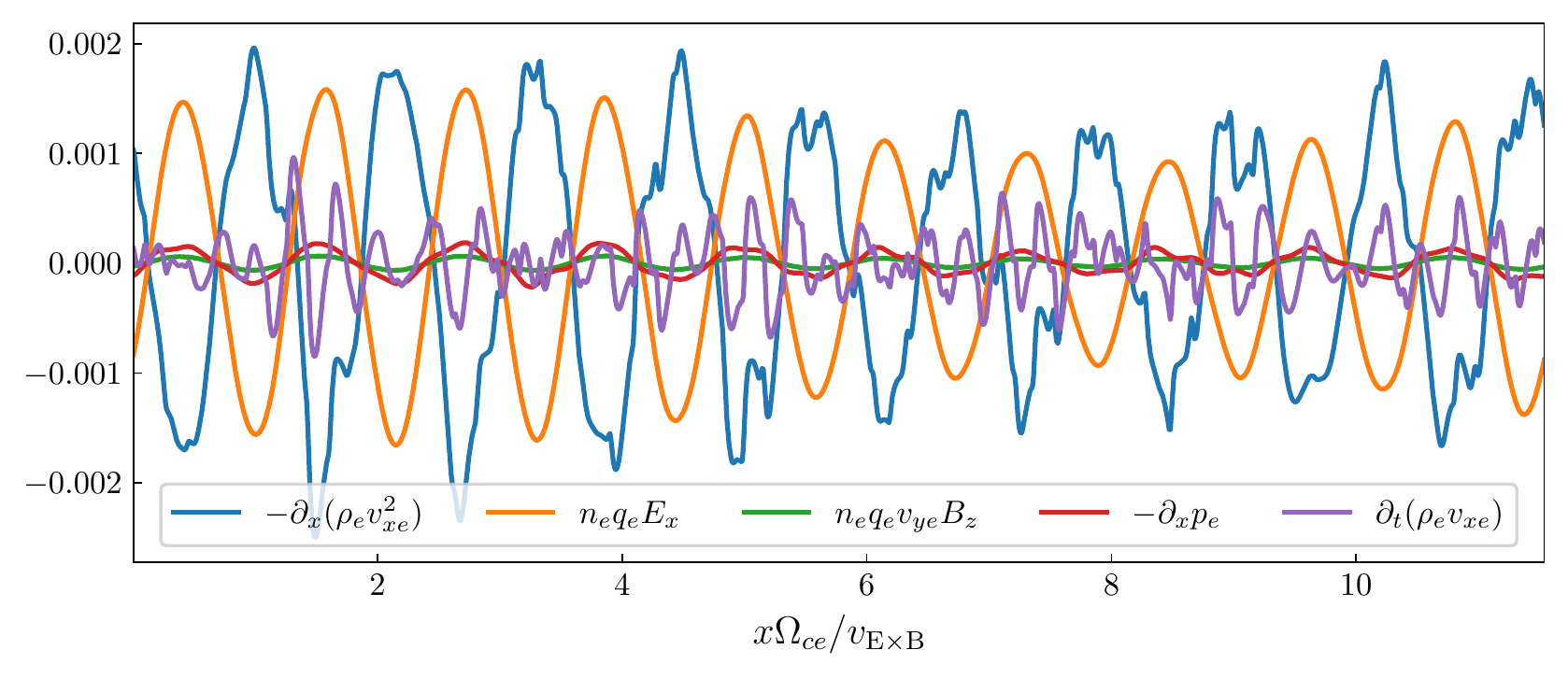}
\par\end{center}%
\end{minipage}%
\begin{minipage}[c]{0.45\textwidth}%
\begin{center}
\includegraphics[width=1\textwidth]{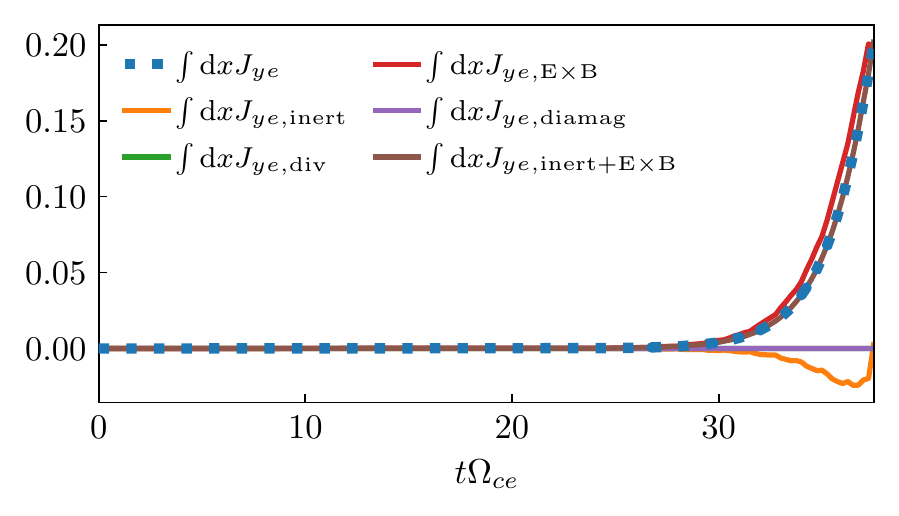}
\par\end{center}%
\end{minipage}
\par\end{centering}
\caption{\label{fig:x_force_balance-and-Jye-decomp}\textbf{Left}: Components
of the electron momentum equation along the $x$ (azimuthal) direction
in the mid-linear stage at $t=25\Omega_{ce}^{-1}$. The magenta term
(the last term in the figure legend) is the
net acceleration and are the summation of the remaining terms. 
\textbf{Right}:
Temporal development of the spatially-integrated axial electron current
$\int_{0}^{L}J_{ye}{\rm d}x$ (dotted blue curve) and its decomposition
(various solid curves). The terms corresponding equation~(\ref{eq:J-decomp}) are $J_{ye}=n_{e}q_{e}v_{ey}$,
$J_{ye,{\rm E\times B}}=n_{e}q_{e}E_{x}/B_{z}$, $J_{ye,{\rm inert}}=-\partial_{t}\left(\rho_{e}v_{xe}\right)/B_{z}$,
$J_{ye,{\rm div}}=-\partial_{x}\left(\rho_{e}v_{xe}^{2}\right)/B_{z}$,
$J_{ye,{\rm diamag}}=-\partial_{x}p_{e}/B_{z}$.}

\end{figure*}

So far, we have shown the development of nonzero $\left\langle n_{e}E_{x}\right\rangle $
and the electron anomalous current. It is useful to further examine
how they are related. We start by examining the electron momentum
equation along the azimuthal direction $x$,
\begin{align}
 & \frac{\partial\left(\rho_{e}v_{xe}\right)}{\partial t}+\frac{\partial p_{e}}{\partial x}+\frac{\partial\left(\rho_{e}v_{xe}^{2}\right)}{\partial x}\nonumber \\
 & =n_{e}q_{e}\left(E_{x}+v_{ey}B_{z}\right),
 \label{eq:azi_mom}
\end{align}
to understand the role of the azimuthal electric force term $n_{e}q_{e}E_{x}$.
The left panel of \Figref{x_force_balance-and-Jye-decomp} shows
the decomposition of this equation in the middle of the linear stage
at $t=25\Omega_{ce}^{-1}$. In this snapshot,
The pressure gradient force (red) is smaller in magnitude and the $V\times B$
force (green) term is negligible.
The $n_{e}q_{e}E_{x}$
term (orange) and the flow divergence term $\partial_{x}\left(\rho v_{xe}^{2}\right)$
(blue) are much larger in magnitude but appear to largely cancel each
other.
The net acceleration, i.e., the time derivative term (magenta), is
a fraction of the $n_{e}q_{e}E_{x}$ and $\partial_{x}\left(\rho v_{xe}^{2}\right)$
terms, indicating the importance of both terms.

Next, we divide this momentum equation by $B_{z}$ and integrate each of the
components of Eq. \ref{eq:azi_mom} along the $x$-coordinate at every time step. This way,
we obtain the time evolution of the net and decomposed currents along
the $y$-direction (axial),
\begin{equation}
\int{\rm d}x\left[n_{e}q_{e}v_{ey}+\frac{n_{e}q_{e}E_{x}}{B_{z}}-\frac{1}{B_{z}}\frac{\partial\left(\rho_{e}v_{xe}\right)}{\partial t}-\frac{1}{B_{z}}\frac{\partial p_{e}}{\partial x}-\frac{1}{B_{z}}\frac{\partial\left(\rho_{e}v_{xe}^{2}\right)}{\partial x}\right].\label{eq:J-decomp}
\end{equation}
The results are shown in the right panel of \Figref{x_force_balance-and-Jye-decomp}.
Here, the dashed curve is the total current $\int n_{e}q_{e}v_{ye}{\rm d}x$.
The current due to the flow divergence
term (green) and the  pressure gradient (magenta) are negligible. The main contribution comes from the
${\rm E}_{x}\times{\rm B}_{z}$ current $\int{\rm d}xn_{e}q_{e}E_{x}/B_{z}$
(red), with small cancellation due to time-derivative inertial term,
$-\int{\rm d}x\partial_{t}\left(\rho_{e}v_{xe}\right)/B_{z}$ (orange).
The sum of the two agrees with the net anomalous current very well.
Therefore, in this simulation, the anomalous electron current is supported
by the ${\rm E\times B}$ flux, consistent with some previous works
\citep{Lafleur2016b} but differs from the conclusions of 
\citep{Janhunen2018c} (the latter observed $\rm E \times B$ fluxes that is not large enough to fully account for the anoamlous transport). A more comprehensive understanding of this
issue would require exhaustive numerical experiments in various parameter
regimes and could be the topic of future work.

\section{\label{sec:Comparing-with-Vlasov}{5-Moment Fluid vs. Fully-Kinetic Simulations Using Experimental Parameters}}

\begin{figure}
\begin{centering}
\begin{minipage}[c]{0.43\columnwidth}%
\begin{center}
\includegraphics[width=1\columnwidth]{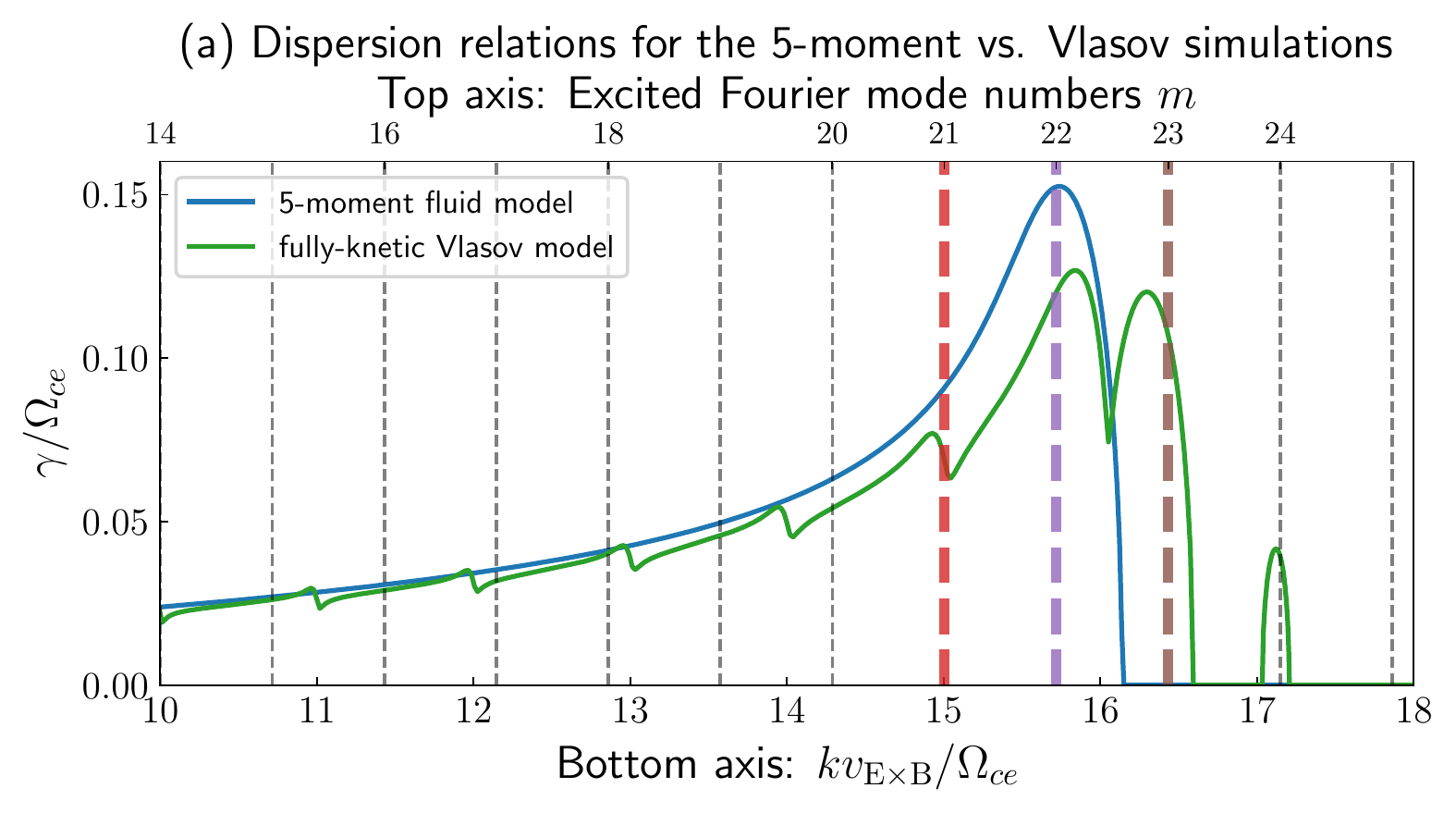}
\par\end{center}
\begin{center}
\includegraphics[width=1\columnwidth]{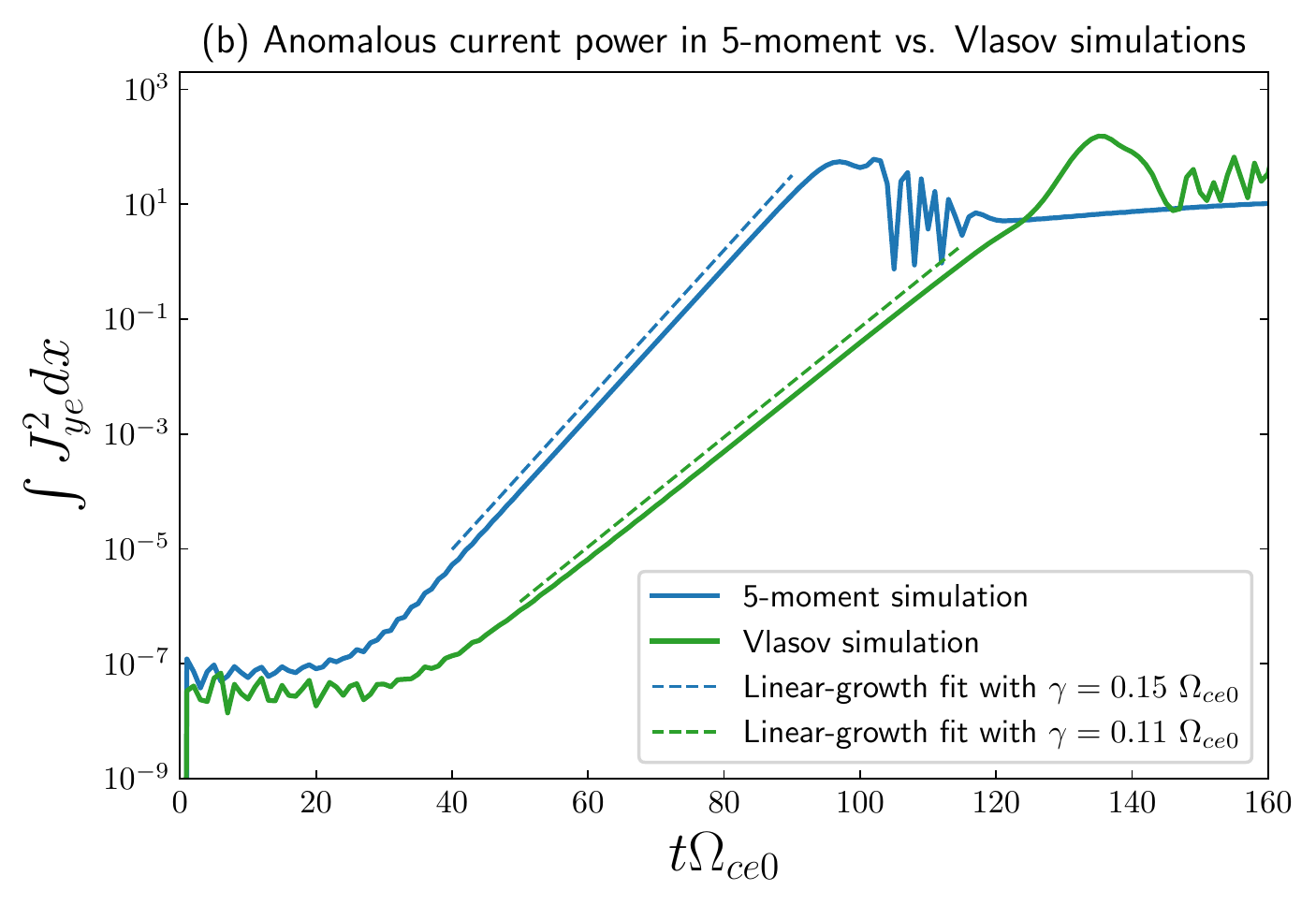}
\par\end{center}%
\end{minipage}%
\begin{minipage}[c]{0.57\textwidth}%
\begin{center}
\includegraphics[width=1\columnwidth]{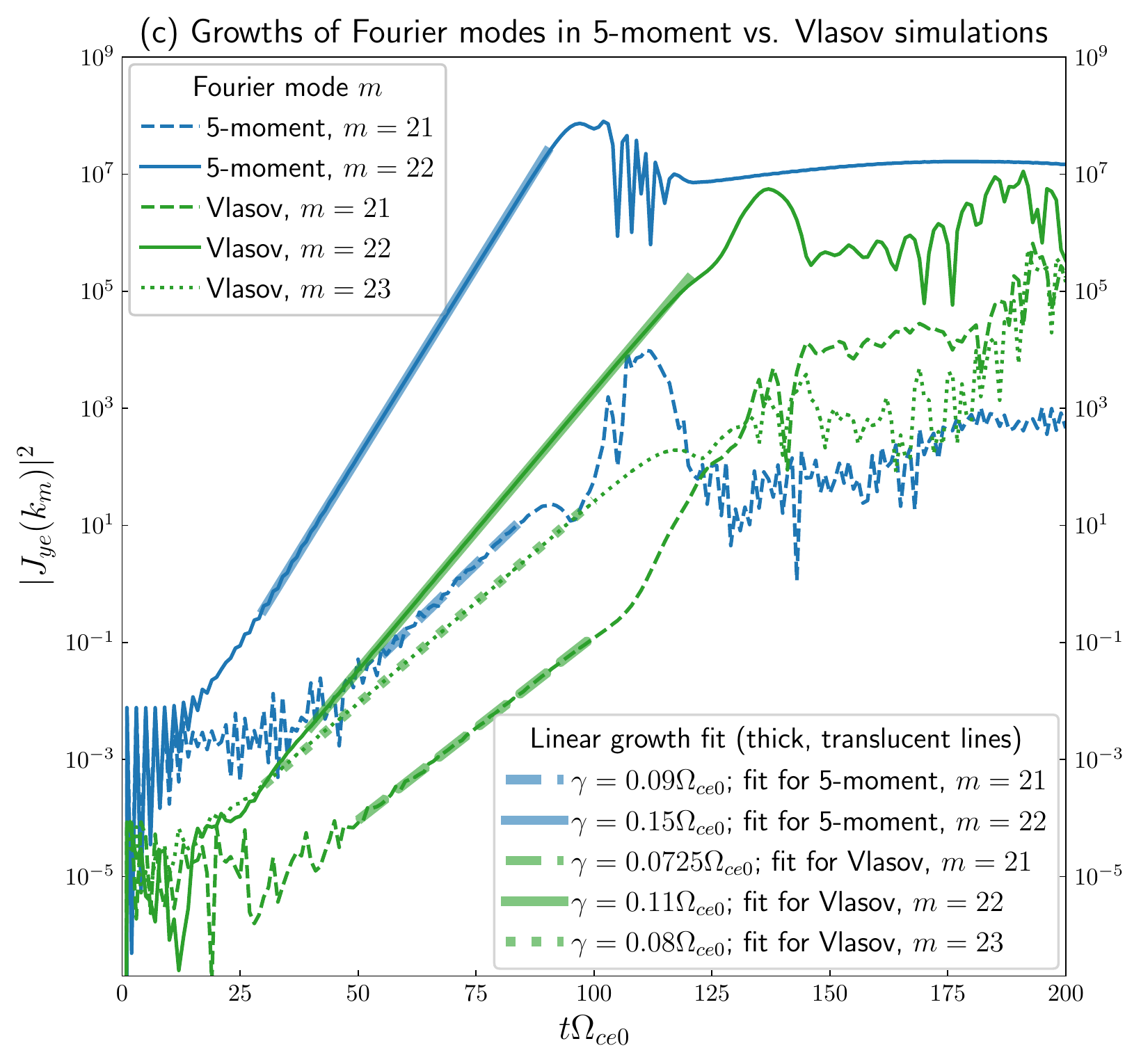}
\par\end{center}%
\end{minipage}\caption{\label{fig:compare-sim-dr-and-growth}{
	Comparison of 5-moment and Vlasov simulations in Section.~\ref{sec:Comparing-with-Vlasov} using parameters $E_0=20~{\rm kV/m}$, $B_0=0.005~{\rm T}$, $m_i/m_e=241074$, $n_0=5\times10^{16}~{\rm m^{3}}$, $T_e=5~{\rm eV}$, and $T_i=0.1~{\rm eV}$. All 5-moment diagnostics are in blue, and all Vlasov diagnostics are in green.
\textbf{Upper left panel (a)}:
Dispersion relations of the ECDI for parameters used in the comparative
5-moment vs. Vlasov simulations. This panel is a zoomed-in version of Figure 1b
with additional vertical dashed lines marking wavenumbers of initial
Fourier modes $k_{m}=2\pi m/L_{x}$, where mode numbers $m=1,2,\cdots,30$.
Particularly, the red, purple, and brown vertical lines denote mode
numbers $m=21$, $22$, and $23$, which are within the range of peak
growths for either model. \textbf{Lower left panel (b)}: Time evoluation
of the integrated electron anomalous current power in the 5-moment
(blue curve) and Vlasov (green curve) simulations. The dashed straight
lines represent the linear-growth fits for the two runs.
\textbf{Right panel} (c): The growth of different Fourier modes in
the 5-moment (blue curves) and Vlasov (green curves) simulations,
and linear-growth fits (thick, translucent straight lines).}}
\par\end{centering}
\end{figure}

In this section, we compare 1D 5-moment and fully-kinetic Vlasov-Poisson
simulations of the instability using parameters relevant to ${\rm E \times B}$ devices.
The goal is to demonstrate
the capability and limitations of the model for realistic, experimental parameters,
particularly in the linear growth and saturation of the
anomalous electron current.
We chose the second case in the linear theory
comparison in Section \ref{subsec:dr-compare} which has
$E_0=20~{\rm kV/m}$, $B_0=0.005~{\rm T}$, $m_i/m_e=241074$, $n_0=5\times10^{16}~{\rm m^{3}}$, $T_e=5~{\rm eV}$, and $T_i=0.1~{\rm eV}$. These parameters give the dispersion relations shown in Figure \ref{fig:dr-compare}b, or the zoomed-in version of Figure \ref{fig:compare-sim-dr-and-growth}a.

The simulation coordinates are consistent with those used in Section \ref{sec:Numerical-Simulations}.
The domain length is set to $L_{x}=4\,{\rm cm}$ and 30
modes in the electron density were initialized at wavenumbers $k_{m}=2\pi m/L_{x}$,
$m=1,2,\cdots30$, with relative density fluctuations $\delta n_{m}/n_{0}=m\times10^{-7}$
so that all modes have equal magnitudes of electric field fluctuations.
These wavenumbers are marked by vertical dashed lines in Figure \ref{fig:compare-sim-dr-and-growth}a.
Particularly, the orange, green, and red vertical lines denote mode
numbers $m=21$, $22$, and $23$, which are within the range of peak
growths.
In these initialized modes, the 5-moment prediction's $m=22$ mode has the greatest growth rate $\gamma\approx0.152\,\Omega_{ce0}$, and the $m=21$ mode has the second greatest $\gamma\approx0.091\,\Omega_{ce0}$.
The kinetic prediction's $m=21$, $22$, and $23$ modes have growth rates $\gamma\approx0.0691\,\Omega_{ce0}$, $\gamma\approx0.120\,\Omega_{ce0}$, and $0.107\,\Omega_{ce0}$, respectively.

For the kinetic simulation, we used the continuum Vlasov-Poisson
module available in Gkeyll v2 (the same code providing the 5-moment
fluid simulation)\citep{Juno2018,Cagas2017}. The kinetic simulation uses a Discontinuous-Galerkin finite-element
scheme and 2nd-order Serendipity bases for spatial discretization and
a 2nd-order Strong-Stability preserving Runge-Kutta (SSP-RK2) scheme
for time integration. The number of spatial cells are $2560$ and $320$,
respectively, in the 5-moment and Vlasov simulations. The kinetic simulation
also uses a square $v_{x}\times v_{y}$ velocity domain of widths
$24\,v_{{\rm th}s}$ and cell numbers $64$ for either species $s=e,i$.

Figure \ref{fig:compare-sim-dr-and-growth}b shows the growth of total spatially integrated
anomalous electron current power in the two simulations. Fitting of the 5-moment
simulation between $40<t\Omega_{ce0}<90$ gives a linear growth rate $\gamma\approx0.15\,\Omega_{ce0}$,
in excellent agreement with the theoretical prediction. Fitting of
the kinetic simulation gives a growth rate $\gamma\approx0.11\,\Omega_{ce0}$,
between the first two fastest growth rates predicted, as they are
close in the first place. It is interesting to note that two simulations
come down to comparable plateau values. Therefore, for these parameters, the 5-moment
simulation gives a qualitatively reasonable prediction of the saturation level for the
total anomalous current in terms of order of magnitude. The agreement
would vary with the parameters like the electron temperature.

As implied by the dispersion relations in Figure \ref{fig:compare-sim-dr-and-growth}a,
the kinetic simulation would allow two comparable modes to develop
at $m=22$ and $23$, while the 5-moment fluid simulation only has
one dominating mode at $m=22$. The simulations confirmed this difference.
Figure \ref{fig:compare-sim-dr-and-growth}c shows the power of different
Fourier components of the anomalous current $J_{ye}$ (for simplicity,
only the dominating modes between $21\leq m\leq 23$ are shown). The
simulations clearly captured the expected linear growth rates, marked
by thick, translucent straight lines, for either model.

\begin{figure}
\begin{centering}
\begin{minipage}[c]{0.5\textwidth}%
\begin{center}
\includegraphics[width=1\columnwidth]{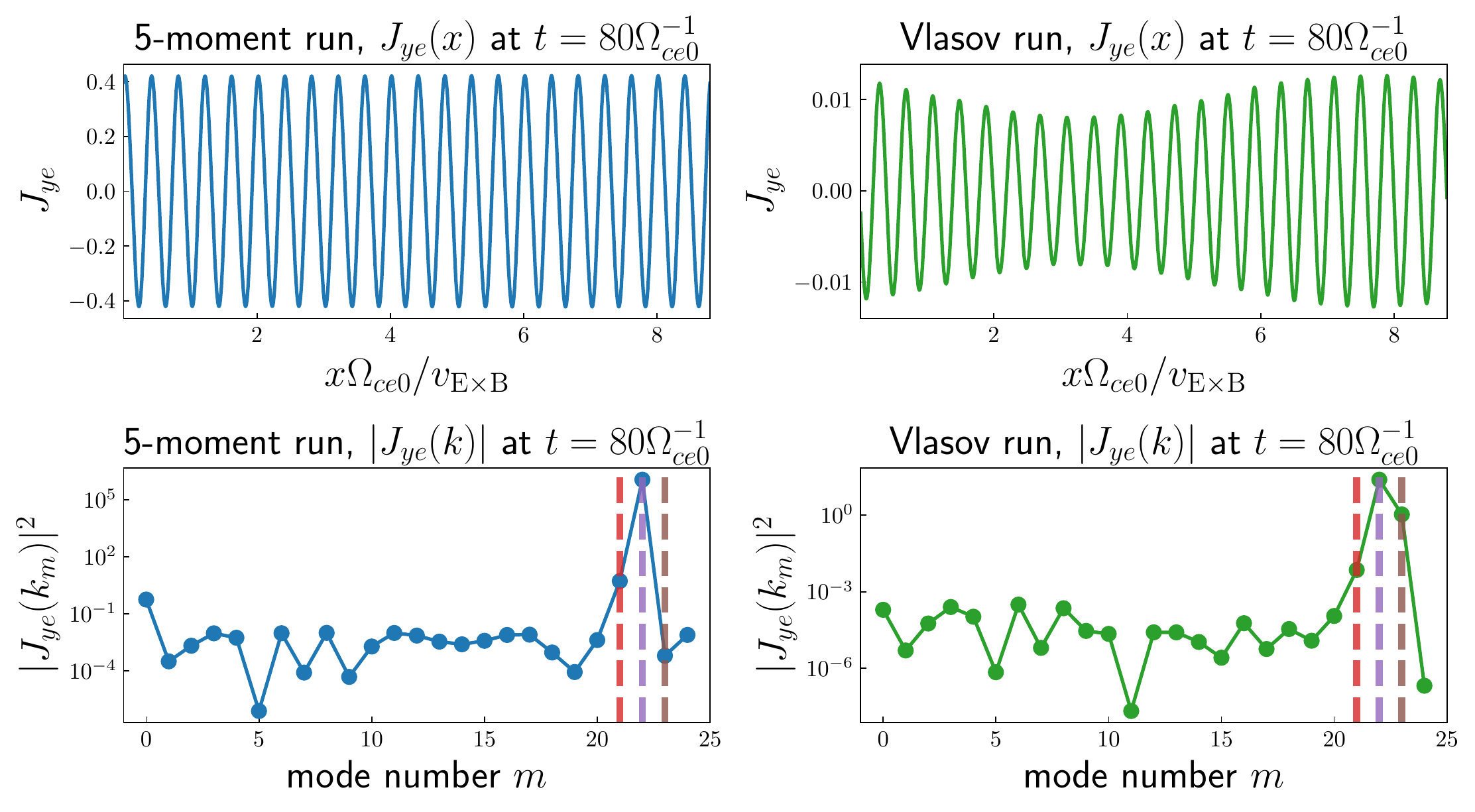}
\par\end{center}%
\end{minipage}{\vrule width 0.2pt}%
\begin{minipage}[c]{0.5\textwidth}%
\begin{center}
\includegraphics[width=1\columnwidth]{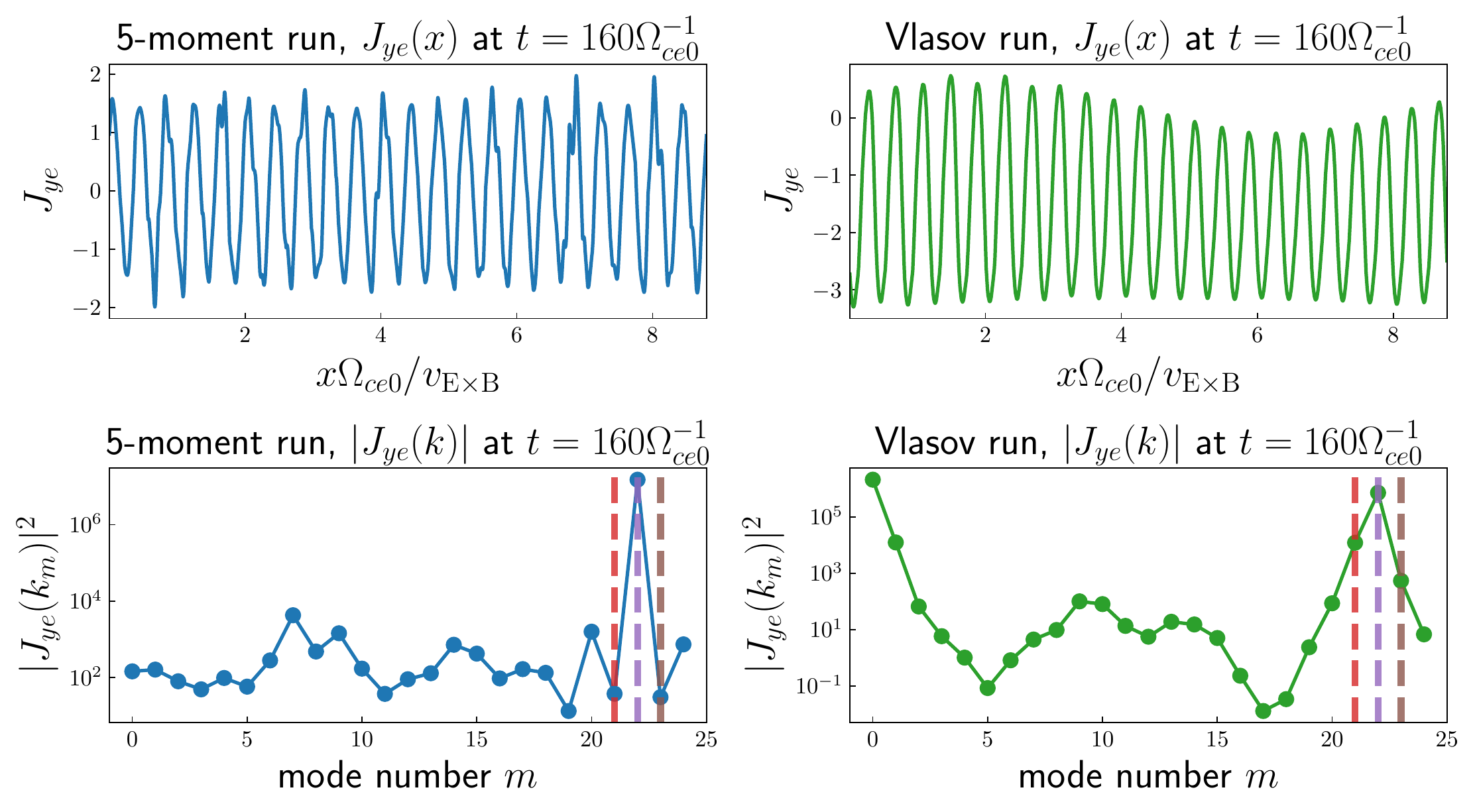}
\par\end{center}%
\end{minipage}
\par\end{centering}
\centering{}\caption{\label{fig:compare-sim-5m-vp-profiles}
	More comparison of 5-moment and Vlasov simulations in Section.~\ref{sec:Comparing-with-Vlasov}.
	{\textit{Left four panel}s:
For a linear stage frame at $t=80\,\Omega_{ce0}^{-1}$, the spatial
profile of the anomalous current in the 5-moment (upper left, blue line) and
Vlasov (upper right, green line) simulations, the power of the first Fourier modes
(lower left, blue line with dots) and Vlasov (lower right, green line with dots) simulations. Consistent with Figure \ref{fig:compare-sim-dr-and-growth}a,
the red, purple, and brown vertical dashed lines denote mode numbers $m=21$,
$22$, and $23$.
 \textit{Right four panels}: The same diagnostics
in the saturation stage at $t=160\,\Omega_{ce0}^{-1}$.}}
\end{figure}

Figure \ref{fig:compare-sim-5m-vp-profiles} shows the configuration
and wavenumber space profiles of the anomalous current during typical
linear (left, at $t=80\,\Omega_{ce0}^{-1}$) and saturation (left,
at $t=160\,\Omega_{ce0}^{-1}$) stages.
In the linear stage, both
the spatial profile and Fourier component powers clearly show the
dominance of the $m=22$ mode in the 5-moment simulation, while the
kinetic simulation shows the overlapping of and competence between the $m=22$
and $m=23$ modes. In the saturation stage, the 5-moment simulation
patterns become very ``spiky'' and remain dominated by the single
$m=22$ mode, while the kinetic simulation develops a broader range
of wave modes, notably at larger wavelengths.

In summary, using realistic experimental parameters and at relatively
low electron temperature, the 5-moment fluid model is capable of capturing
the growth of anomalous current roughly in the correct regime. The
5-moment model itself clearly lacks the broad electron harmonics excited
in the fully-kinetic simulation, but the total saturated anomalous
current seems to be a good indicator of the kinetic values for lower temperatures. Again,
such agreement relies on the parameter regime and becomes less satisfactory
as the temperature rises and higher cyclotron harmonics become important. On the other hand, the multifluid high-moment model may indeedcapture
more cyclotron harmonics by including higher velocity
moments in future studies.

\section{\label{sec:Discussions-and-Conclusions}Discussions and Conclusions}

The ECDI due to the electron ${\rm E\times B}$ drift in a cross-field
setup is an important research topic actively studied in the HET community
and has been drawing increased attention from the space physics
community as well. Traditional ECDI studies often rely on fully kinetic models. Models based on the fluid or hybrid
description are often thought to require additional collision models
and adjustable parameters to correctly capture produce anomalous transport.
In this paper, we show detailed theoretical proof
and numerical evidence how this instability develops in a collisionless
two-fluid plasma, and leads to enhanced axial electron anomalous transport.

In the 5-moment model, only the lowest-order
electron resonance is captured, and the coupling between a Doppler
shifted hybrid wave associated with this resonance, and an ion-acoustic-like
wave, leads to the development of ECDI. Compared to the fully kinetic
theory featured by a highly quantized nature of the unstable modes due to
higher-order electron resonances, the 5-moment gives reasonable prediction
of the fastest-growing mode in terms of both wavelength and growth
rate, when the plasma temperature is low. The prediction gets worse in comparison to fully kinetic descriptions as the plasma temperature increases.

As indicated by the secondary unstable branch when using the 10-moment
model, we may capture a more accurate dispersion relation, including
the discrete patterns noted in the kinetic description, by including higher fluid moments. Due to the
very low cost of these fluid moment models, this provides a promising
new approach for future modeling of HETs and other space physics phenomena where cross-field
instabilities are important.

We presented preliminary comparison against fully-kinetic simulations using realistic, experimental parameters, which confirmed the model's ability to predict the growth and saturation of anomalous current roughly in the correct regime at relatively low electron temperatures.
The comparison also shows the model's inherent limitation of not being able to excite higher harmonics and broader wave modes, which may be partially overcome by including higher velocity moments (like the heat-flux tensor) in the fluid equations. 
Finally, it should be noted that, the focus of this work is on the fundamental properties and scaling of the dispersion relation and the development of anomalous transport without additional collisions.
The work performed here may be extended in the future to use the 10-moment model, and possibly even higher-order moment fluid models, with improved plasma closure relations based on physical constraints~\citep{Hammett1990,AllmannRahn2018,Ng2018a,Ng2020,boccelli2020collisionless,Boccelli2022} or data-driven approaches~\citep{CHENG2022108538}.

\begin{acknowledgments}
This work was supported by the Air Force Office of Scientific Research under grant number FA9550-15-1-0193. The work of Bhuvana Srinivasan was supported by the National Science Foundation under grant number PHY-1847905. The work of Ammar Hakim was also partially supported via DOE contract DE-AC02-09CH11466 for the Princeton Plasma Physics Laboratory. Liang Wang thanks Manaure Francisquez for suggestions on continuum Vlasov-Poisson simulations.
\end{acknowledgments}

\bibliographystyle{plainnat}
\bibliography{ecdi-5m}

\end{document}